\documentclass[11pt,fleqn]{article}       
\usepackage{geometry}
\usepackage[T1]{fontenc}
\usepackage[utf8]{inputenc}
\usepackage{authblk}
\usepackage{mathrsfs}
\usepackage[font=small,labelfont=bf]{caption}
\usepackage{graphicx}
\usepackage{multirow}
\usepackage{amsmath,amssymb,amsfonts}
\usepackage{verbatim}
\usepackage{bm}
\usepackage{color}
\usepackage{ulem}
\usepackage{mathtools}
\usepackage{cite}
\usepackage{colortbl}
\definecolor{light-gray}{gray}{0.85}
\usepackage[pdftex,colorlinks,pdfpagelabels]{hyperref}
\hypersetup{
   bookmarksnumbered,
   citecolor={blue},
   linkcolor={blue},
   urlcolor={blue},
   filecolor={blue}
} 
\newcommand{\eVdist}{\kern-0.06em}

\newcommand{\gev}{\:\text{Ge\eVdist V}}

\newcommand{\re}{\:\text{Re}\,}
\newcommand{\im}{\:\text{Im}\,}
\newcommand{\tev}{\:\text{Te\eVdist V}}
\newcommand{\SU}[1]{\ensuremath{\mathrm{SU}(#1)}}

\newcommand{\Th}{\ensuremath{T_{\scalebox{.7}{H}}}}
\newcommand{\Thcaption}{\ensuremath{T_{\protect\scalebox{.7}{H}}}}
\newcommand{\Thmin}{\ensuremath{T_{\scalebox{.7}{H},0}}}
\newcommand{\Tlmin}{\ensuremath{T_{\scalebox{.7}{L},0}}}
\newcommand{\Tlcaption}{\ensuremath{T_{\protect\scalebox{.7}{L}}}}
\newcommand{\Thlcaption}{\ensuremath{T_{\protect\scalebox{.7}{H,L}}}}
\newcommand{\Tlmincaption}{\ensuremath{T_{\protect\scalebox{.7}{L},0}}}
\newcommand{\Tl}{\ensuremath{T_{\scalebox{.7}{L}}}}
\newcommand{\barTl}{\ensuremath{\overline{T}_{\scalebox{.7}{L}}}}
\newcommand{\WH}{\ensuremath{\mathcal{W}_{\scalebox{.7}{H}}}}

\newcommand{\WHH}{\ensuremath{\mathcal{W}_{\scalebox{.7}{HH}}}}

\newcommand{\Kl}{\ensuremath{K_{\scalebox{.7}{L}}}}
\newcommand{\Kllbar}{\ensuremath{K_{\scalebox{.7}{$\mathrm{L}\overline{\mathrm{L}}$}}}}

\newcommand{\AddrMichigan}{%
\textit{Leinweber Center for Theoretical Physics, Department of Physics, University of Michigan, Ann Arbor, MI 48109, USA}
}
\newcommand{\AddrStockholm}{
\textit{The Oskar Klein Centre for Cosmoparticle Physics, Department of Physics, Stockholm University, Alba Nova, 10691 Stockholm, Sweden}
}

\usepackage{fancyhdr}
 
\fancypagestyle{plain}{%
    \fancyhead[R]{LCTP-19-02}
    
}
\date{}
\title{\Large\bf Deriving the Inflaton in Compactified M-theory with a \\ De Sitter Vacuum}
\author[1]{Gordon Kane}
\author[2]{Martin Wolfgang Winkler\thanks{martin.winkler@su.se}}
\affil[1]{\AddrMichigan}
\affil[2]{\AddrStockholm}

\begin{document}
\maketitle
\vspace*{0mm}
\begin{abstract}
Compactifying M-theory on a manifold of $G_2$ holonomy gives a UV complete 4D theory. It is supersymmetric, with soft supersymmetry breaking via gaugino condensation that simultaneously stabilizes all moduli and generates a hierarchy between the Planck and the Fermi scale. It generically has gauge matter, chiral fermions, and several other important features of our world. Here we show that the theory also contains a successful inflaton, which is a linear combination of moduli closely aligned with the overall volume modulus of the compactified $G_2$ manifold. The scheme does not rely on ad hoc assumptions, but derives from an effective quantum theory of gravity. Inflation arises near an inflection point in the potential which can be deformed into a local minimum. This implies that a de Sitter vacuum can occur in the moduli potential even without uplifting. Generically present charged hidden sector matter generates a de Sitter vacuum as well.
\end{abstract}
\clearpage

\tableofcontents
\section{Introduction}
Countless papers have suggested particles or fields that can lead to an inflating universe. Most have used ad hoc mechanisms without identifying a physical origin – what is the inflaton? Such bottom-up descriptions, furthermore, rely on strong hidden assumptions on the theory of quantum gravity. 
More thorough proposals have identified the inflaton as part of a string theory construction in which the ultraviolet (UV) physics can be addressed. In this case, the inflaton arises in a theory that itself satisfies major consistency conditions and tests. The theory should also connect with the Standard Models of particle physics and cosmology.  Ideally, its properties would uniquely determine the nature of the inflaton.

In this work, we focus on M-theory compactified spontaneously on a manifold of $G_2$ holonomy. The resulting quantum theory is UV complete and describes gravity plus the Standard Model plus Higgs physics. When its hidden sector matter is included it has a de Sitter vacuum~\cite{Acharya:2007rc}. It stabilizes all the moduli, and is supersymmetric with supersymmetry softly broken via gluino condensation and gravity  mediated~\cite{Acharya:2007rc}. It produces a hierarchy of scales, and has quarks and leptons interacting via Yang-Mills forces. It generically has radiative electroweak symmetry breaking, and correctly anticipated the ratio of the Higgs boson mass to the $Z$ mass~\cite{Kane:2011kj}. It also solves the strong CP problem~\cite{Acharya:2010zx}. 

In this theory, a particular linear combination of moduli, that which describes the volume of the compactified region, generates inflation. By means of K\"ahler geometry, we will prove that a tachyonic instability develops if the inflaton is not `volume modulus-like’. In contrast to related proposals in type II string theory~\cite{Linde:2007jn,Badziak:2008yg,Conlon:2008cj}, volume modulus inflation on $G_2$ does not rely on uplifting or higher order corrections to the K\"ahler potential. This follows from the smaller curvature on the associated K\"ahler submanifold.

Besides being intuitively a likely inflaton, the volume modulus also resolves a notorious problem of string inflation: the energy density injected by inflation can destabilize moduli fields and decompactify the extra dimensions. Prominent moduli stabilization schemes including KKLT~\cite{Kachru:2003aw}, the large volume scenario~\cite{Balasubramanian:2005zx} and K\"ahler uplifting~\cite{Balasubramanian:2004uy,Westphal:2006tn} share the property that the volume modulus participates in supersymmetry breaking. Its stability is threatened once the Hubble scale of inflation $H$ exceeds $m_{3/2}$~\cite{Buchmuller:2004xr,Kallosh:2004yh,Buchmuller:2015oma}. In contrast, the volume modulus of the compactified $G_2$ manifold drives inflation in the models we will discuss. Thereby, the inflationary energy density stabilizes the system and $H \gg m_{3/2}$ is realized. The supersymmetry breaking fields - light moduli and mesons of a strong hidden sector gauge theory - receive stabilizing Hubble mass terms on the inflationary trajectory. 

Inflation takes place close to an inflection point in the potential and lasts for 100-200 e-foldings. If we impose the observational constraints on the spectral index, we can predict the tensor-to-scalar ratio $r\sim 10^{-6}$.  It is unlikely that other observables will directly probe the nature of the inflaton. However, inflation emerges as piece of a theory which also implies low energy supersymmetry with a gravitino mass $m_{3/2} \lesssim 100\tev$ and a specific pattern of superpartner masses. Gauginos are at the TeV scale and observable at LHC. Furthermore, a matter dominated cosmological history is predicted. In a sense, all aspects and tests of the theory are also tests of the nature of its inflaton, although technically they may not be closely related.

Less is known about $G_2$ manifolds than about Calabi-Yau manifolds.  This is being at least partially remedied via a 4-year, 9 million \$ study sponsored by the Simons Foundation started in 2017, focusing on $G_2$ manifolds.  Remarkably, the above successes were achieved without detailed knowledge of the properties of the manifolds.

\section{De Sitter Vacua in $G_2$ Compactifications}\label{sec:g2vacua}

\subsection{The Moduli Sector}
We study M-theory compactifications on a flux-free $G_2$-manifold. The size and the shape of the manifold is controlled by moduli $T_i$. In our convention, the imaginary parts of the $T_i$ are axion fields.\footnote{$T_i$ in this work corresponds to $i z_i$ defined in~\cite{Acharya:2007rc}.} A consistent set of K\"ahler potentials is of the form~\cite{Beasley:2002db,Acharya:2005ez}
\begin{equation}
 K=-3\log\left(4\pi^{1/3}\mathcal{V}\right)\,,
\end{equation}
where $\mathcal{V}$ denotes the volume of the manifold in units of the the eleven-dimensional Planck length. Since the volume must be a homogeneous function of the $\re T_i$ of degree 7/3, the following simple ansatz has been suggested~\cite{Acharya:2005ez}
\begin{equation}\label{eq:MKahler}
 K=-\log\left[\frac{\pi}{2} \prod\limits_i (\overline{T}_i+T_i)^{a_i}\right]\,,\qquad \sum\limits_{i} a_i =7\,,
\end{equation}
which corresponds to $\mathcal{V}= \prod_i (\re T_i)^{a_i/3}$. We will drop the factor $\pi/2$ in the following since it merely leads to an overall $\mathcal{O}(1)$ factor in the potential not relevant for this discussion. A realistic vacuum structure with stabilized moduli is realized through hidden sector strong dynamics such as gaugino condensation. The resulting theory generically has massless quarks and leptons, and Yang-Mills forces~\cite{Acharya:2007rc}, and it has generic electroweak symmetry breaking, and no strong CP problem~\cite{Acharya:2010zx}.

We consider one or several hidden sector $\SU{N}$ gauge theories. These may include massless quark states $Q$, $\overline{Q}$ transforming in the $N$ and $\overline{N}$ representations. Each hidden sector induces a non-perturbative superpotential due to gaugino condensation~\cite{Seiberg:1993vc,Seiberg:1994bz}
\begin{equation}
 W = A \,\det{\left(Q\overline{Q}\right)}^{-\frac{1}{N-N_f}} \, \exp\left({-\frac{2\pi\,f}{N-N_f}}\right)\,,
\end{equation}
where $N_f$ denotes the number of quark flavors. The coefficient $A$ is calculable, but depends on the RG-scheme as well as threshold corrections to the gauge coupling. The gauge kinetic function $f$ is a linear combination of the moduli~\cite{Lukas:2003dn},
\begin{equation}
 f= c_i T_i\,,
\end{equation}
with integer coefficients $c_i$. We now turn to the construction of de Sitter vacua with broken supersymmetry.

\subsection{Constraints on de Sitter Vacua}\label{sec:ConstraintsdS}

In this section we introduce some tools of K\"ahler geometry which can be used to derive generic constraints on de Sitter vacua in supergravity~\cite{GomezReino:2006dk}. The same framework also applies to inflationary solutions (see e.g.~\cite{Badziak:2008yg}) and will later be employed to identify the inflaton field. In order to fix our notation, we introduce the ($F$-term part) of the scalar potential in supergravity
\begin{equation}
 V= e^G\left( G^i G_{i} - 3\right)\,,
\end{equation}
with the function $G=K + \log|W|^2$. The subscript $i$ indicates differentiation with respect to the complex scalar field $\phi_i$. Indices can be raised and lowered by the K\"ahler metric $K_{i\bar{j}}$ and its inverse $K^{\bar{i}j}$. Extrema of the potential satisfy the stationary conditions $V_i=0$ which can be expressed as
\begin{equation}
 e^G (G_i + G^j \nabla_i G_j) + G_i V =0\,,
\end{equation}
where we introduced the K\"ahler covariant derivatives $\nabla_i$. The mass matrix at stationary points derives from the second derivatives of the potential~\cite{Covi:2008ea}, 
\begin{align}
V_{i \bar j} &=  e^{G} \left(G_{i \bar j}  +  \nabla_i G_k \nabla_{\bar j}  G^k
-  R_{i \bar j m \bar n} \, G^m  G^{\bar n} \right) + \left(G_{i \bar j}  -  G_i G_{\bar j}\right) V \,,
\label{eq:Vibarj}\\
V_{i j} &= e^{G} \left(2 \nabla_i G_j+  G^k \nabla_i \! \nabla_j G_k \right)
+ \left(\nabla_i G_j -  G_i G_j \right) V \,,
\end{align}
where $R_{i \bar j m \bar n}$ denotes the Riemann tensor of the K\"ahler manifold. (Meta)stable vacua are obtained if the mass matrix is positive semi-definite. A weaker necessary condition requires the submatrix $V_{i \bar j}$ to be positive semi-definite. All complex scalars orthogonal to the sgoldstino may acquire a large mass from the superpotential. In addition, the above mass matrix contains the standard soft terms relevant e.g.\ for the superfields of the visible sector. 

Stability constraints apply in particular to the sgoldstino direction which does not receive a supersymmetric mass. Via appropriate field redefinitions, we can set all derivatives of $G$ to zero, except from one which we choose to be $G_n$. The curvature scalar of the one-dimensional submanifold associated with the sgoldstino is defined as
\begin{equation}\label{eq:curvaturescalar}
 R_n=
 \frac{K_{nn\bar{n}\bar{n}}}{K_{n\bar{n}}^2}-\frac{K_{nn\bar{n}}\,K_{n\bar{n}\bar{n}}}{K_{n\bar{n}}^3}\,.
\end{equation} 
From the necessary condition, it follows that $V_{n\bar{n}}\geq 0$ and, hence,
\begin{equation}\label{eq:curvaturecondition0}
 e^G\, (2- 3 R_n) - V R_n \geq 0\,.
\end{equation}
For a tiny positive vacuum energy as in the observed universe, the constraint essentially becomes~\cite{GomezReino:2006dk}
\begin{equation}\label{eq:curvaturecondition}
 R_n < \frac{2}{3}\,.
\end{equation}
This condition restricts the K\"ahler potential of the field responsible for supersymmetry breakdown. Indeed, it invalidates some early attempts to incorporate supersymmetry breaking in string theory. For the dilaton $S$ in heterotic string theory, one can e.g.\ derive the curvature scalar $R_S=2$ from its K\"ahler potential $K=-\log(\overline{S}+S)$. The scenario of dilaton-dominated supersymmetry breaking~\cite{Kaplunovsky:1993rd} is, hence, inconsistent with the presence of a de Sitter minimum~\cite{Brustein:2000mq,GomezReino:2006dk}. K\"ahler potentials of the no-scale type $K= -3\log (\overline{T}+T)$, with $T$ denoting an overall K\"ahler modulus, feature $R_T = 2/3$. In this case~\eqref{eq:curvaturecondition} is marginally violated. Corrections to the K\"ahler potential and/ or subdominant $F$ or $D$-terms from other fields may then reconcile $T$-dominated supersymmetry breaking with the bound. Examples of this type include the large volume scenario~\cite{Balasubramanian:2005zx} as well as K\"ahler uplifting~\cite{Balasubramanian:2004uy,Westphal:2006tn}. 

A less constrained possibility to realize de Sitter vacua consists in the supersymmetry breaking by a hidden sector matter field. 
Hidden sector matter is present in compactified M-theory.  When it is included using the approach of Seiberg~\cite{Seiberg:1994bz}, it generically leads to a de Sitter vacuum.
The identification of the goldstino with the meson of a hidden sector strong gauge group allows for a natural explanation of the smallness of the supersymmetry breaking scale (and correspondingly the weak scale) through dimensional transmutation. The simple canonical K\"ahler potential, for instance, yields a vanishing curvature scalar consistent with~\eqref{eq:curvaturecondition}. Matter supersymmetry breaking is also employed in KKLT modulus stabilization~\cite{Kachru:2003aw} with $F$-term uplifting~\cite{Lebedev:2006qq} and in heterotic string models~\cite{Lowen:2008fm}.

We note, however, that in $G_2$ compactifications of M-theory, de Sitter vacua can arise even if the hidden sector matter decouples. As we show in section~\ref{sec:modularinflation}, the $G_2$ K\"ahler potential~\eqref{eq:MKahler} features linear combinations of moduli with curvature scalar as small as 2/7.  In contrast to the previously mentioned string theory examples, condition~\eqref{eq:curvaturecondition} can hence be satisfied even in the absence of corrections to the K\"ahler potential. The modular inflation models we discuss in section~\ref{sec:modularinflation} are of this type. We will show that, by a small parameter deformation, the inflationary plateau can be turned into a metastable de Sitter minimum.

Let us also briefly allude to the controversy on the existence of de Sitter vacua in string/ M-theory~\cite{Obied:2018sgi}. It is known that de Sitter vacua do not arise in the classical limit of string/ M-theory~\cite{Maldacena:2000mw}. This, however, leaves the possibility to realize de Sitter vacua at the quantum level. Indeed, in the $G_2$ compactification we describe, the scalar potential is generated by quantum effects. The quantum nature is at the heart of the proposal and tied to the origin of physical scales.

\subsection{Minimal Example of Modulus Stabilization}\label{sec:singlemodulus}

We describe the basic mechanism of modulus stabilization in $G_2$-compactifications leaning on~\cite{Acharya:2007rc}.\footnote{Some differences occur since~\cite{Acharya:2007rc} mostly focused on the case of two hidden sector gauge groups with equal gauge kinetic functions, while we will consider more general cases.} Some key features are illustrated within a simple one-modulus example. Since the single-modulus case faces cosmological problems which can be resolved in a setup with two or more moduli, we will later introduce a two-moduli example and comment on the generalization to many moduli.

The minimal example\footnote{Due to the absence of a constant term in the superpotential, a single gaugino condensate would give rise to a runaway potential} of modulus stabilization in $G_2$-compactifications invokes two hidden sector gauge groups $\SU{N_1+1}$, $\SU{N_2}$ with gauge kinetic functions
\begin{equation}
 f_1=f_2=T\,.
\end{equation}
The $\SU{N_1+1}$ gauge theory shall contain one pair of massless quarks $Q$, $\overline{Q}$ transforming in the fundamental and anti-fundamental representation of $\SU{N_1+1}$. When the $\SU{N_1+1}$ condenses, the quarks form an effective meson field $\phi=\sqrt{2Q\overline{Q}}$. Taking $\SU{N_2}$ to be matter-free, the superpotential and K\"ahler potential read
\begin{align}\label{eq:onemod}
 W &= A_1 \,\phi^{-\frac{2}{N_1}} \, e^{-\frac{2\pi T}{N_1}} + A_2 \, e^{-\frac{2\pi T}{N_2}} \,,\nonumber\\
 K &= -7 \log\left(\overline{T}+T\right) + \overline{\phi}\phi\,,
\end{align}
We negelected the volume dependence of the matter K\"ahler potential which does qualitatively not affect the modulus stabilization~\cite{Acharya:2008hi}. The scalar potential including the modulus and meson field is
\begin{equation}\label{eq:potentialonemodulus}
 V= e^G\left( G^T G_T + G^{\phi} G_{\phi} - 3\right)\,.
\end{equation}
The scalar mass spectrum contains two CP even and two CP odd (axion) states which are linear combinations of $\re T$, $|\phi|$ and $\im T$, $\arg \phi$ respectively. We will denote the CP even and odd mass eigenstates by $s_{1,2}$ and $\varphi_{1,2}$ respectively. The scalar potential is invariant under the shift
\begin{equation}
T\rightarrow T + i \frac{N_2}{N_1-N_2}\,\Delta\,,  \qquad\phi \rightarrow e^{i\pi\Delta} \phi\,.
\end{equation}
This can easily be seen from the fact that the superpotential merely picks up an overall phase under this transformation. The light axion 
\begin{equation}
\varphi_1 \propto N_2 \im T + \pi(N_1-N_2) \arg \phi 
\end{equation}
is, hence, massless which makes it a natural candidate for the QCD axion~\cite{Acharya:2010zx}. The remaining axionic degree of freedom receives a periodic potential which has an extremum at the origin of field space. Without loss of generality, we require $\text{sign}(A_1/A_2)=-1$ such that the extremum is a minimum.\footnote{If this condition is not satisfied, the relative sign of $A_1$ and $A_2$ can be inverted through field redefinition.} This allows us to set $\im T=\arg \phi=0$ when discussing the stabilization of the CP even scalars.

We now want to prove that this setup allows for the presence of a (local) de Sitter minimum consistent with observation. For practical purposes, we can neglect the tiny cosmological constant and require the presence of a Minkowski minimum with broken supersymmetry. There is generically no supersymmetric minimum at finite field values. Since the negative sign of $A_1/A_2$ is required for axion stabilization, a solution to $G_T=0$ only exists if $N_2>N_1$. With this constraint imposed, there is no simultaneous solution to $G_{\phi}=0$ with positive $|\phi|$. However, a minimum $(T_0,\phi_0)$ with broken supersymmetry may occur close to the field value $T_{\text{susy}}$ at which $G_T$ vanishes. This is because the modulus mass term at $T_{\text{susy}}$ dominates over the linear term which drives it away from this point.
Given a minimum with a small shift $\delta T = T_{\text{susy}}-T_0$, we can expand
\begin{equation}\label{eq:GT}
  G_{T} = G_{\bar{T}} = - (G_{TT}+G_{T\bar{T}}) \,\delta T\,.
\end{equation}
Here and in the following, all terms are evaluated at the minimum if not stated otherwise. Since $T_0$, $\phi_0$, $\delta T$ are real, there is no need to distinguish between $G_T$ and $G_{\bar{T}}$. In order to determine the shift, we insert~\eqref{eq:GT} into the minimization condition $V_T=0$ and keep terms up to linear order in $\delta T$.
Notice that all derivatives of $G$ with respect to purely holomorphic or purely antiholomorphic variables are of zeroth order in $T_0^{-1}$. We find
\begin{equation}\label{eq:shift}
 \delta T = \frac{G_{\phi T} G_{\bar{\phi}}}{G_{TT}K^{T\bar{T}}G_{\bar{T}\bar{T}}} + \mathcal{O}\left(T_0^{-4}\right)\,.
\end{equation}
The leading contribution to the shift is $\delta{T}=\mathcal{O}(T_0^{-2})$. This justifies our expansion in $\delta T$. In the next step, we want to determine the location of the minimum. As an additional constraint, we require a vanishing vacuum energy. In order to provide simple analytic results, we will perform a volume expansion which is equivalent to an expansion in $T_0^{-1}$. We include terms up to $\mathcal{O}(T_0^{-1})$. Notice that, at this order, the modulus minimum satisfies $T_0 = T_{\text{susy}}$. We, nevertheless, have to keep track of the shift carefully since it may appear in a product with the inverse K\"ahler metric which compensates its suppression. The conditions $V_T=V_\phi=V=0$ lead to the set of equations at order $T_0^{-1}$
\begin{equation}\label{eq:minimum}
 G_{T}=0\,,\qquad
 G_{\phi\phi}+1-\frac{G_{\phi T}^2}{G_{TT}}=0\,,\qquad
 G_{\phi}= \sqrt{3}\,.
\end{equation}
The solutions for the modulus and meson minimum read
\begin{equation}
 \phi_0=\frac{\sqrt{3}}{2}\,,\qquad  T_0=\frac{14}{\pi}\, \frac{N_2}{3(N_2-N_1)-8}\,.
\end{equation}
Notice that a minimum only exists for $N_2 \geq N_1 +3$. On the other hand $N_2-N_1\lesssim 10$ since the non-perturbative terms in the superpotential would otherwise exceed unity. The equations~\eqref{eq:minimum} fix one additional parameter which can be taken to be the ratio $A_1/A_2$. We find
\begin{equation}\label{eq:parameter}
 \frac{A_1}{A_2} =-\frac{N_1}{N_2}\left(\frac{3}{4}\right)^{\frac{1}{N_1}} \exp\left[\frac{28}{N_1}\frac{N_2-N_1}{3\,(N_2-N_1)-8}\right]\,.
\end{equation}
A suppressed vacuum energy can be realized on those $G_2$ manifolds which fulfill the above constraint\footnote{More accurately, the exact version of the above approximate constraint.} with acceptable precision.
We now turn to the details of supersymmetry breaking. The gravitino mass is defined as
\begin{equation}
 m_{3/2}= |e^{G/2}|_{T_0,\phi_0}\,.
\end{equation}
Throughout this work, $m_{3/2}$ refers to the gravitino mass in the vacuum of the theory. We will later also introduce the gravitino mass during inflation, but will clearly indicate the latter by an additional superscript $I$. Within the analytic approximation, the gravitino mass determined from~\eqref{eq:minimum} and~\eqref{eq:parameter} is
\begin{equation}
 m_{3/2} \simeq |A_1|\,\frac{e^{3/8}\pi^{7/2}}{48 N_1}\,\left(\frac{3 N_2- 3 N_1 -8}{7 N_2}\right)^{7/2}\,\exp\left[-\frac{N_2}{N_1}\,\frac{28}{ 3(N_2-N_1)-8}\right]\,.
\end{equation}
Up to the overall prefactor, the gravitino mass is fixed by the rank of the hidden sector gauge groups. A hierarchy between the Planck scale and the supersymmetry breaking scale naturally arises from the dimensional transmutation. If we require a gravitino mass close to the electroweak scale, this singles out the choice $N_2 = N_1 + 4$. While this particular result only holds for the single modulus case, similar relations between the gravitino mass and the hidden sector gauge theories can be established in realistic systems with many moduli~\cite{Acharya:2007rc}.\footnote{In realistic $G_2$ compactifications, the gauge kinetic function is set by a linear combination of many moduli. We can effectively account for this by modifying the gauge kinetic function to $f= \mathcal{O}(10-100) \,T$ in the one-modulus example. In this case, the preferred value of $N_2-N_1$ changes to 3 in agreement with~\cite{Acharya:2007rc}.} In order to determine the pattern of supersymmetry breaking we evaluate the $F$-terms which are defined in the usual way,
\begin{equation}
 F^i=e^{G/2} K^{i\bar{j}} G_{\bar{j}}\,.
\end{equation}
From~\eqref{eq:GT} and~\eqref{eq:shift}, we derive 
\begin{equation}
 |F^T| \simeq \frac{2 N_2}{\pi(N_2-N_1)}\,m_{3/2} \,,\qquad
 |F^{\phi}| \simeq \sqrt{3} \,m_{3/2}
\end{equation}
at leading order. The meson provides the dominant source of supersymmetry breaking as can be seen by comparing the canonically normalized $F$-terms
\begin{equation}
 \frac{\left|F^T \sqrt{K_{\bar{T}T}}\right|}{\left|F^{\phi}\right|} \simeq \frac{3 N_2-3 N_1-8}{2\sqrt{21}(N_2-N_1)}\,.
\end{equation}
This has important implications for the mediation of supersymmetry breaking to the visible sector. Since gravity-mediated gaugino masses only arise from moduli $F$-terms, they are suppressed against the gravitino and sfermion masses. We refer to~\cite{Acharya:2008zi} for details.

As stated earlier, the modulus and the meson are subject to mixing. However, the mixing angle is suppressed by $T_0$, and the heavy CP even and odd mass eigenstates $s_2$ and $\varphi_2$ are modulus-like. Since their mass is dominated by the supersymmetric contribution $m_{\bar{T}T}$, they are nearly degenerate with
\begin{equation}
 m_{s_2} \simeq m_{\varphi_2} \simeq e^{G/2}\,\sqrt{\frac{G_{TT}K^{\bar{T}T}G_{\bar{T}\bar{T}}}{K_{\bar{T}T}}} \simeq 
 \frac{56}{N_1}\,\frac{3 N_2^2-3 N_1^2 -8 N_1}{(3 N_2 -3 N_1-8)^2}\,m_{3/2}\,.
\end{equation}
The meson-like axion $\varphi_1$ is massless due to the shift symmetry. Since the meson is the dominant source of supersymmetry breaking, the supertrace of masses in the meson multiplet must approximately cancel. This implies
\begin{equation}
 m_{s_1} \simeq 2\, m_{3/2}\,.
\end{equation}
The scalar potential vanishes towards large modulus field values. Hence, the minimum ($T_0,\phi_0$) is only protected by a finite barrier. We first keep the meson fixed and estimate its height in a leading order volume expansion.\footnote{We also assumed $N_{1,2}\gg N_2-N_1$ when estimating the barrier height.} Then, we allow the meson to float, in order to account for a decrease of the barrier in the mixed modulus-meson direction. Numerically, we find that the shifting meson generically reduces the barrier height by another factor $\sim T_0^{-1}$. Our final estimate thus reads
\begin{equation}\label{eq:barrierpotential}
 V_{\text{barrier}} \simeq \frac{16\pi^2 T_0}{7 e^2 N_1^2}\,m_{3/2}^2\,.
\end{equation}
The prefactor in front of the gravitino mass is of order unity. Notice that the above expression is multiplied by two powers of the Planck mass which is set to unity in our convention.

For illustration, we now turn to an explicit numerical example. We choose the following parameter set
\begin{equation}\label{eq:benchmarkparameter}
 N_1=8\,,\qquad N_2=12\,,\qquad A_1=0.0001\,.
\end{equation}
The prefactor $A_2$ is fixed by requiring a vanishing vacuum energy. Numerically, we find 
\begin{equation}
 A_1/A_2=-20.9\,,
\end{equation}
in good agreement with the analytic approximation~\eqref{eq:parameter}. We list the resulting minimum, particle masses, supersymmetry breaking pattern and barrier height in table~\ref{tab:spectrum1}. The numerical results are compared with the analytic expressions provided in this section. The approximations are valid to within a few per cent precision. Only for $m_{3/2}$ the error is larger due to its exponential dependence on the modulus minimum.

\begin{table}[h]
\begin{center}
 \begin{tabular}{|cc|cccccccc|}
 \hline
&&&&&&&&&\\[-4mm]
 $T_0$ & $\phi_0$ & $m_{3/2}$ & $\!m_{\varphi_1}\!$ & $m_{\varphi_2}$ & $m_{s_1}$ & $m_{s_2}$&  $F^T$&  $F^\phi$ &$V_{\text{barrier}}$\\[1mm]
 \hline\hline
$12.9$ & $0.85$ & $57\tev$ & $0$ & $77.1\, m_{3/2}$ & $1.98\, m_{3/2}$ & $75.4\, m_{3/2}$&  $1.98 \,m_{3/2}$&  $1.72 \,m_{3/2}$ &$0.5\,m_{3/2}^2$\\
$13.4$ & $0.87$ & $33\tev$ & $0$ & $77\, m_{3/2}$ & $2\, m_{3/2}$ & $77\, m_{3/2}$ &  $1.91\, m_{3/2}$ &  $1.73\, m_{3/2}$ &$0.6\,m_{3/2}^2$\\
 \hline  
 \end{tabular} 
\end{center}
\caption{Location of the minimum, mass spectrum, $F$-terms and height of the potential barrier for the parameter choice~\eqref{eq:benchmarkparameter}. The upper and lower line correspond to exact numerical result and analytic approximation respectively.}
\label{tab:spectrum1}
\end{table}

\begin{figure}[t]
\begin{center}   
 \includegraphics[height=5.3cm]{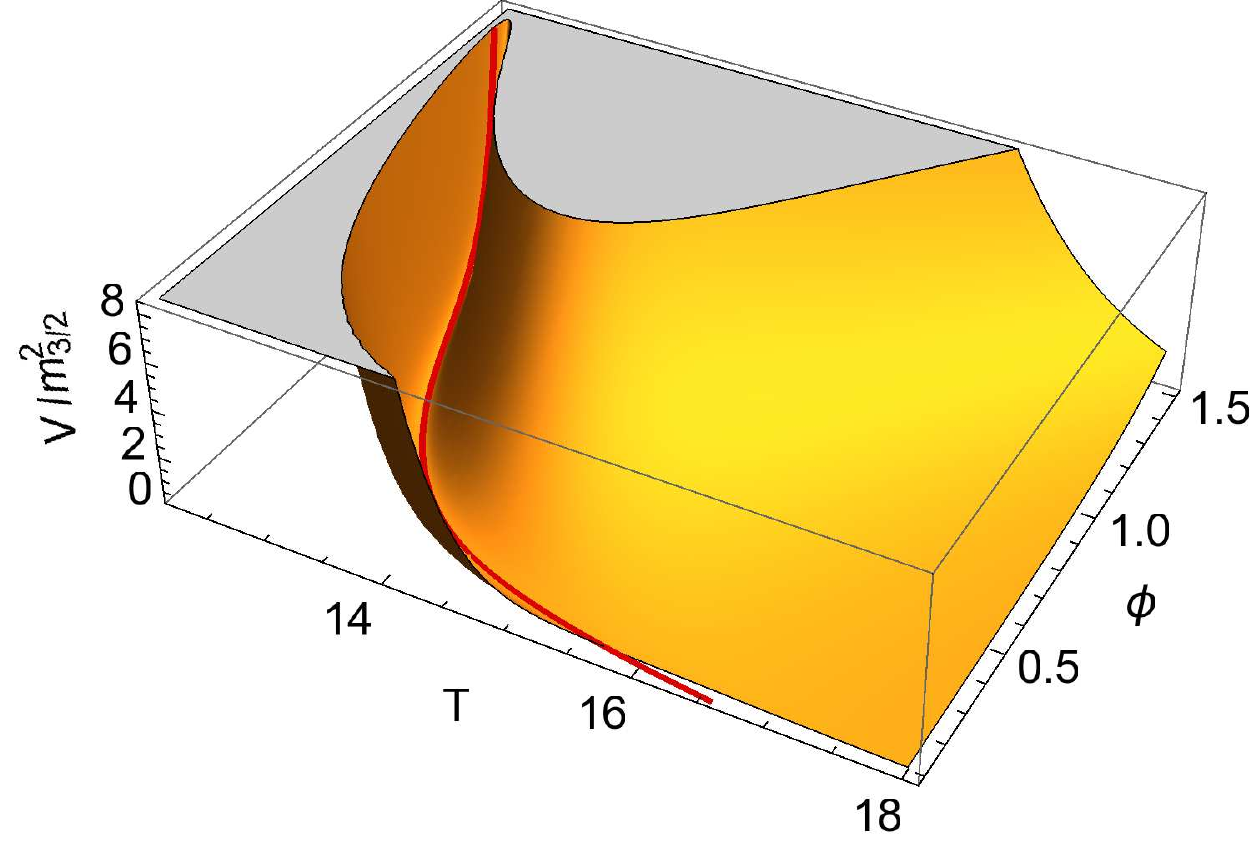}\hspace{6mm}
 \includegraphics[height=4.7cm]{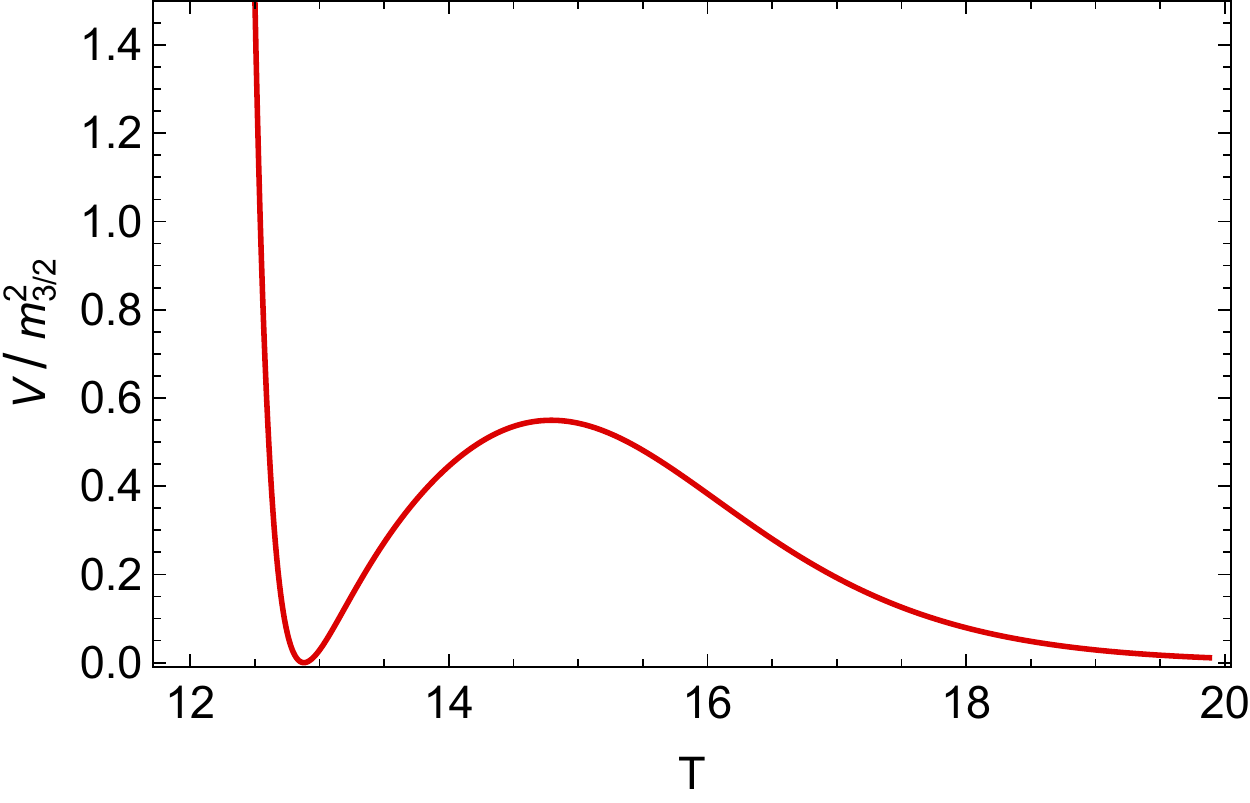} 
\end{center}
\caption{The left panel shows the scalar potential (in Planck units) in modulus and meson direction rescaled by $m_{3/2}^2$. A local minimum with broken supersymmetry is located at $T_0=12.9$, $\phi_0=0.85$. The field direction with the shallowest potential barrier is indicated by the red line. In the right panel, the potential along this direction is shown.}
\label{fig:modulus1}
\end{figure}

The scalar potential in the modulus-meson plane is depicted in figure~\ref{fig:modulus1}. Also shown is the potential along the `most shallow' mixed modulus-meson direction. The latter was determined by minimizing the potential in meson direction for each value of $T$.

\subsection{Generalization to Several Moduli}\label{sec:severalmoduli}

Realistic $G_2$ manifolds must contain the full MSSM spectrum with its $\mathcal{O}(100)$ couplings. They will generically feature a large number of moduli and non-perturbative terms in the superpotential. The low energy phenomenology, however, mostly depends on the lightest modulus. In this sense, the mass spectrum derived in the previous section is realistic, once $T$ is identified with the lightest modulus. However, in the early universe, high energy scales are accessed. This implies that, for cosmology, the heavier moduli do actually matter. We will later see that inflation in M-theory relies on large mass hierarchies in the moduli sector. In order to motivate their existence, we now introduce an example with two moduli $T_{1,2}$.

One linear combination of moduli $\Tl$ plays the role of the light modulus as in the previous section. It participates (subdominantly) in supersymmetry breaking and its mass is tied to the gravitino mass. The orthogonal linear combination $\Th$ can, however, be decoupled through a large supersymmetric mass term from the superpotential. In order to be explicit, we will identify 
\begin{equation}
 \Th = \frac{T_1 + T_2}{2}\,,\qquad \Tl = \frac{T_1 - T_2}{2}\,.
\end{equation}
The superpotential is assumed to be of the form
\begin{equation}
 W = \mathcal{W}(\Th) + w(\Th,\Tl)\,,
\end{equation}
The part $\mathcal{W}$ only depends on $\Th$ and provides the large supersymmetric mass for the heavy linear combination. The part $w$ is responsible for supersymmetry breaking and its magnitude is controlled by the (much smaller) gravitino mass. We require that $\Th$ is stabilized supersymmetrically at a high mass scale. For this we impose that the high energy theory defined by $\mathcal{W}$ has a supersymmetric Minkowski minimum, i.e.
\begin{equation}\label{eq:globalsusy}
\mathcal{W}=\WH = 0\,,
\end{equation}
where the subscript $\text{H}$ indicates differentiation with respect to $\Th$. The above condition has to be fulfilled at the minimum which we denote by $\Thmin$. It ensures that $\Th$ can be integrated out at the superfield level. The mass of the heavy modulus is given as
\begin{equation}\label{eq:heavymodmass}
m_{\Th}\simeq \left| e^{K/2}\:\WHH\:\left(\frac{1}{4K_{1\bar{1}}}+\frac{1}{4K_{2\bar{2}}}\right)\right|
\end{equation}
with $K_{i\bar{i}}$ denoting the entries of the K\"ahler metric in the original field basis. Since $m_{\Th}$ is unrelated to the gravitino mass, it can be parametrically enhanced against the light modulus mass. The construction of a Minkowski minimum for $\Tl$ with softly broken supersymmetry proceeds analogously to the one-modulus case. 

As an example we consider five hidden sector gauge groups $\SU{N_1+1}$ and $\SU{N_i}$ ($i=2,\dots 5$) with gauge kinetic functions
\begin{equation}
 f_{1,2}= 2\,T_1 + T_2\,,\qquad f_{3,4,5}= T_1+T_2\,.
\end{equation}
The $\SU{N_1+1}$ shall again contain one pair of massless quarks $Q$, $\overline{Q}$ forming the meson $\phi=\sqrt{2Q\overline{Q}}$. The remaining gauge theories are taken to be matter-free. Super- and K\"ahler potential take the form
\begin{align}\label{eq:twomod}
 W &= \underbrace{A_1 \,\phi^{-\frac{2}{N_1}} \, e^{-\frac{2\pi f_1}{N_1}} + A_2 \, e^{-\frac{2\pi f_2}{N_2}}}_{w}+ \underbrace{A_3 \, e^{-\frac{2\pi f_3}{N_3}}+ A_4 \, e^{-\frac{2\pi f_4}{N_4}}+ A_5 \, e^{-\frac{2\pi f_5}{N_5}}}_{\mathcal{W}} \,,\nonumber\\[2mm]
 K &= - \log\left(\overline{T}_1+T_1\right)-6 \log\left(\overline{T}_2+T_2\right) + \overline{\phi}\phi\,.
\end{align}
We have assumed
\begin{equation}\label{eq:hierarchy}
 |A_1 \, e^{-\frac{2\pi f_1}{N_1}}|,\,|A_2 \, e^{-\frac{2\pi f_2}{N_2}}| \;\;\ll\;\; |A_3 \, e^{-\frac{2\pi f_3}{N_3}}|,\,|A_4 \, e^{-\frac{2\pi f_4}{N_4}}| ,\,|A_5 \, e^{-\frac{2\pi f_5}{N_5}}|\,,
\end{equation}
such that the first two gaugino condensates contribute to $w$, the last three to $\mathcal{W}$. In order to obtain a supersymmetric minimum with vanishing vacuum energy for the heavy modulus, we impose~\eqref{eq:globalsusy}, which fixes one of the coefficients,
\begin{equation}\label{eq:A5condition}
A_5= -A_3 \left(\frac{A_3}{A_4}\frac{\mathcal{N}_{53}}{\mathcal{N}_{45}}\right)^{\frac{\mathcal{N}_{53}}{\mathcal{N}_{34}}} - A_4 \left(\frac{A_3}{A_4}\frac{\mathcal{N}_{53}}{\mathcal{N}_{45}}\right)^{\frac{\mathcal{N}_{54}}{\mathcal{N}_{34}}}\quad\text{with}\;\;\mathcal{N}_{ij}=\frac{1}{N_i}-\frac{1}{N_j}\,.
\end{equation}
The location of the heavy modulus minimum is found to be
\begin{equation}\label{eq:THmin}
 \Thmin= \frac{\log \left(\frac{A_3}{A_4}\frac{\mathcal{N}_{53}}{\mathcal{N}_{45}}\right)}{4\pi \mathcal{N}_{34} }\,.
\end{equation}
We can now integrate out $\Th$ at the superfield level. In the limit where $\Th$ becomes infinitely heavy, the low energy effective theory is defined by the superpotential $W_\text{eff}=w$ (evaluated at $\Th=\Thmin$) and the K\"ahler potential
\begin{equation}
K_\text{eff} = - \log\left(2\Thmin+\barTl+\Tl\right)-6 \log\left(2\Thmin-\barTl-\Tl\right) + \overline{\phi}\phi\,.
\end{equation}
The effective theory resembles the one-modulus example of the previous section. 
At leading order in the volume expansion, the minimum with softly broken supersymmetry derives from the set of equations~\eqref{eq:minimum} with $T$ replaced by $\Tl$. We find
\begin{equation}\label{eq:TLmin}
 \phi_0=\frac{\sqrt{3}}{2}\,,\qquad  \Tlmin=-\frac{4 \Kl\Tlmin}{\pi}\, \frac{N_2}{3(N_2-N_1)-8}\,,
\end{equation}
where we wrote the equation for $\Tlmin$ in implicit form. In contrast to the single modulus example, values $N_2<N_1+3$ may now be realized since the derivative of the K\"ahler potential $\Kl$ can take both signs. In order for the vacuum energy to vanish, the coefficients $A_{1,2}$ need to fulfill the relation 
\begin{equation}\label{eq:A1condition}
 \frac{A_1}{A_2} = -\frac{N_1}{N_2}\left(\frac{3}{4}\right)^{\frac{1}{N_1}} e^{2\pi(3T_{\text{H,0}}+T_{\text{L,0}})\mathcal{N}_{12}}
\end{equation}
with $\Thmin$ and $\Tlmin$ taken from~\eqref{eq:THmin} and~\eqref{eq:TLmin}. Again, we neglected higher orders in the inverse volume. 
In analogy with section~\ref{sec:singlemodulus}, one can show that the meson provides the dominant source of supersymmetry breaking. The spectrum of scalar fields now contains three CP even states $s_{1,2,3}$ and three CP odd states $\varphi_{1,2,3}$, for which the following mass pattern occurs
\begin{align}\label{eq:massestimate2}
 m_{s_3}&\simeq m_{\Th}\,\quad m_{s_2}\simeq m_{\Tl}= \mathcal{O}\left(\frac{m_{3/2}}{\Kllbar}\right)\,,\quad m_{s_1}=\mathcal{O}\left(m_{3/2}\right)\,,\nonumber\\ 
 m_{\varphi_3}&\simeq m_{\Th}\,\quad m_{\varphi_2}\simeq m_{\Tl}= \mathcal{O}\left(\frac{m_{3/2}}{\Kllbar}\right)\,,\quad m_{\varphi_1}=\mathcal{O}\left(m_{3/2}\sqrt{\frac{m_{\Tl}}{m_{\Th}}}\right)\,,
\end{align}
The heavy states $s_3,\;\varphi_3$ with their mass determined from~\eqref{eq:heavymodmass} are the two degrees of freedom contained in $\Th$. The lighter states are composed of $\Tl$ and $\phi$. They exhibit a similar spectrum as in the single modulus example (section~\ref{sec:singlemodulus}). However, once a finite $m_{\Th}$ is considered, the effective super- and K\"ahler potential receive corrections which are suppressed by inverse powers of $m_{\Th}$. These corrections break the axionic shift symmetry which was present in the one-modulus case. As a result, a non-vanishing mass of the light axion appears. The latter can no longer be identified with the QCD axion. An unbroken shift symmetry can, however, easily be reestablished, once the framework is generalized to include several light moduli.

In order to provide a numerical example, we pick the following hidden sector gauge theories
\begin{equation}\label{eq:twomodulibench}
 A_1=A_3 = 1\,,\;\; A_4=-0.445 \,,\;\;N_1=8 \,,\;\;N_2=10 \,,\;\;N_3=11 \,,\;\;N_4=13 \,,\;\;N_5=15\,.
\end{equation}
The (exact numerical version of the) conditions~\eqref{eq:A5condition} and~\eqref{eq:A1condition} then fixes $A_2=-0.0306$, $A_5=0.0754$. One may wonder, whether the two-moduli example introduces additional tuning compared to the one-modulus case, since two of the $A_i$ are now fixed in order to realize a vanishing cosmological constant. However, deviations from~\eqref{eq:A5condition} and~\eqref{eq:A1condition} can compensate without spoiling the moduli stabilization.\footnote{In the low energy theory, such deviations would manifest as a constant in the superpotential which is acceptable as long as the latter is suppressed against the other superpotential terms.}  Effectively, there is still only a single condition which must be fulfilled to the precision to which the vacuum energy cancels. In table~\ref{tab:spectrum2} we provide the location of the minimum and the resulting mass spectrum for the choice~\eqref{eq:twomodulibench}. 
\begin{table}[h]
\begin{center}
 \begin{tabular}{|ccc|ccccccc|}
 \hline
&&&&&&&&&\\[-4mm]
 $\Thmin\!$ & $\!\Tlmin$ & $\phi_0$ & $m_{3/2}$ & $\!m_{\varphi_1}\!$ & $m_{\varphi_2}$ & $m_{\varphi_3}$ & $m_{s_1}$ & $m_{s_2}$&  $m_{s_3}$\\[1mm]
 \hline\hline
$9.5$ & $\!\!-3.9$ & $0.78$ & $82$ & $2.4$ & $1.4\times 10^3$ & $3.3\times 10^6$&  $148$&  $1.2\times 10^3$ &$3.3\times 10^6$\\
 \hline  
 \end{tabular} 
\end{center}
\caption{Minimum and mass spectrum for the parameter set~\eqref{eq:twomodulibench}. In the original basis, the minimum is located at $T_{1,0}=5.6$, $T_{2,0}=13.4$. All masses are given in $\text{TeV}$.}
\label{tab:spectrum2}
\end{table}
An important observation is that large mass hierarchies -- in this example a factor of $\mathcal{O}(10^3)$ -- can indeed be realized in the moduli sector. The origin of such hierarchies lies in the dimensional transmutation of the hidden sector gauge theories. A larger modulus mass is linked to a higher gaugino condensation scale, originating from a gauge group of higher rank or larger initial gauge coupling.

\begin{figure}[htp]
\begin{center}   
 \includegraphics[height=7cm]{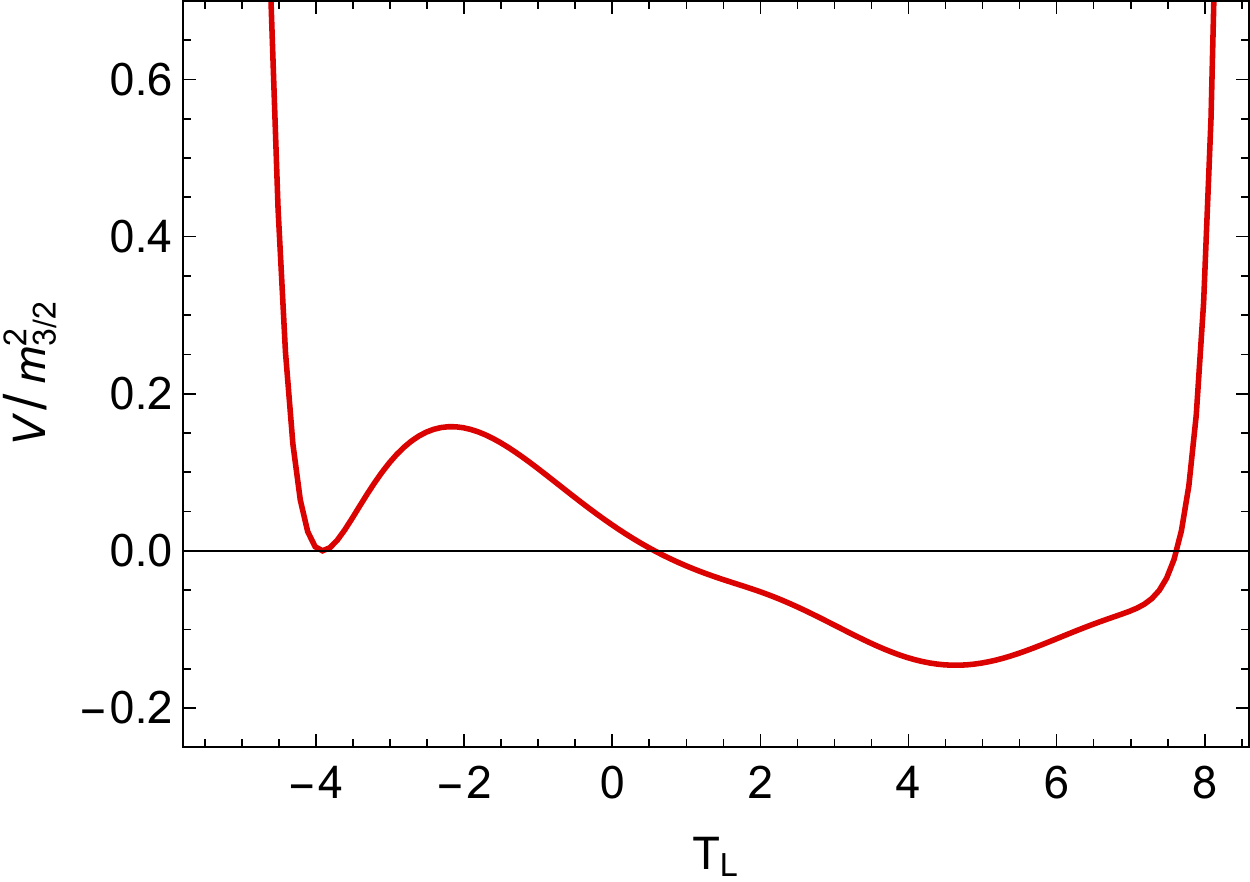}
\end{center}
\caption{Scalar potential along the $\Tlcaption$-direction. The remaining fields were set to their $\Tlcaption$-dependent minima (see text). The Minkowski minimum with softly broken supersymmetry is located at $\Tlmincaption=-3.9$.}
\label{fig:2modulus}
\end{figure}

In figure~\ref{fig:2modulus}, we depict the scalar potential along the light modulus direction. For each value of $\Tl$ we have minimized the potential along the orthogonal field directions. The Minkowski minimum is protected by a potential barrier, in this case against a deeper minimum with negative vacuum energy at $\Tl=4.6$. Similar as in the single modulus example, the barrier height is controlled by the gravitino mass. Numerically, we find $V_{\text{barrier}}= 0.2\,m_{3/2}^2$. The potential rises steeply once $\Tl$ approaches the pole in the K\"ahler metric at $\Tl=\Th$ (corresponding to $T_2=0$). The supergravity approximation breaks down close to the pole which is, however, located sufficiently far away from the Minkowski minimum we are interested in. Of course, we need to require that the cosmological history places the universe in the right vacuum. But once settled there, tunneling to the deeper vacuum does not occur on cosmological time scales as we verified with the formalism~\cite{Coleman:1977py}.

The example of this section can straightforwardly be generalized to incorporate many moduli and hidden sector matter fields. A subset of fields may receive a supersymmetric mass term and decouple from the low energy effective theory. The remaining light degrees of freedom are stabilized by supersymmetry breaking in the same way as $\Tl$ and $\phi$. Indeed, it was shown in~\cite{Acharya:2007rc} that an arbitrary number of light moduli can be fixed through the sum of two gaugino condensates in complete analogy to the examples discussed in this work. 

\section{Modulus (De-)Stabilization During Inflation?}

As was shown in the previous section, the lightest modulus is only protected by a barrier whose seize is controlled by the gravitino mass. There is danger that, during inflation, the large potential energy lifts the modulus over the barrier and destabilizes the extra dimensions. We will show that in the single modulus case, indeed, the bound $H<m_{3/2}$ on the Hubble scale during inflation arises. This constraint was previously pointed out in the context of KKLT modulus stabilization~\cite{Kallosh:2004yh} (the analogous constraint from temperature effects had been derived in~\cite{Buchmuller:2004xr}) and later generalized to the large volume scenario and the K\"ahler uplifting scheme~\cite{Buchmuller:2015oma}. 
The constraint for the single modulus case would leave us with the undesirable choice of either coping with ultra-low scale inflation or of giving up supersymmetry as a solution to the hierarchy problem.\footnote{Another option may consist in fine-tuning several gaugino condensates in order to increase the potential barrier as in models with strong moduli stabilization~\cite{Kallosh:2004yh,Dudas:2012wi}.} As another problematic consequence, supersymmetry breaking would then generically induce large soft terms into the inflation sector which tend to spoil the flatness of the inflaton potential. Fortunately, we will be able to demonstrate that the bound on $H$ does not apply to more realistic examples with several moduli. The crucial point is that in the multi-field case, the modulus which stabilizes the overall volume of the compactified manifold and the modulus participating in supersymmetry breaking in the vacuum are generically distinct fields.

\subsection{Single Modulus Case}

\begin{figure}[h]
\begin{center}   
\includegraphics[height=6cm]{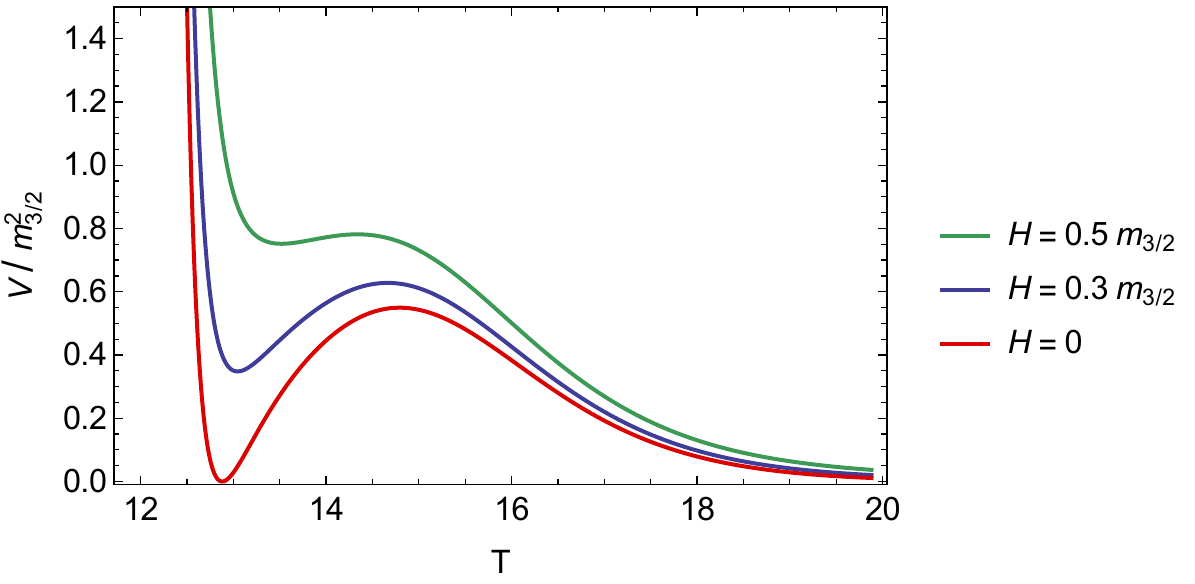} 
\end{center}
\caption{Scalar potential in modulus direction for different choices of the Hubble scale. For each value of $T$, the potential in the meson direction was minimized. Remaining parameters are chosen as in figure~\ref{fig:modulus1}.}
\label{fig:modulus1inf}
\end{figure}

We will now augment the single modulus example by an inflation sector. The latter consists of further moduli or hidden sector matter fields which we denote by $\rho_\alpha$. In order to allow for an analytic discussion of modulus destabilization we shall make some simplifying assumptions. Specifically, we take superpotential and K\"ahler potential to be separable into modulus and inflaton parts,
\begin{equation}
 W = w(T,\phi) + \mathscr{W}(\rho_\alpha) \,,\qquad
 K = k(\overline{T},T,\overline{\phi},\phi) + \mathscr{K}(\overline{\rho}_\alpha,\rho_\alpha)\,.
\end{equation}
 The modulus superpotential $w$ and K\"ahler potential $k$ are defined as in~\eqref{eq:onemod}. As an example inflaton sector, we consider the class of models with a stabilizer field defined in~\cite{Kallosh:2010xz}. These feature
\begin{equation}\label{eq:simple_inflation}
 \mathscr{W}=\mathscr{K}= \mathscr{K}_\alpha=0
\end{equation}
along the inflationary trajectory.\footnote{In this section, we neglect the backreaction of the modulus sector on the inflaton potential. This is justified since, for the moment, we are interested in the stabilization of the modulus during inflation and not in the distinct question, whether the backreaction spoils the flatness of the inflaton potential.}
For now, we focus on modulus destabilization during inflation. Whether this particular inflation model can be realized in M-theory does not matter at this point. In fact, we merely impose the conditions~\eqref{eq:simple_inflation} for convenience since they lead to particularly simple analytic expressions. The important element, which appears universally, is the $e^K\propto (\overline{T}+T)^{-7}$ factor which multiplies all terms in the scalar potential. The latter reads
\begin{equation}
V  = V_{\text{mod}} +  \frac{e^{|\phi|^2}}{(\overline{T}+T)^7} W^{\alpha}W_{\alpha} \,,
\end{equation}
where $V_{\text{mod}}$ coincides with the scalar potential without the inflaton as defined in~\eqref{eq:potentialonemodulus}. The second term on the right hand side sets the energy scale of inflation. It displaces the modulus and the meson. Once the inflationary energy reaches the height of the potential barrier defined in~\eqref{eq:barrierpotential}, the minimum in modulus direction gets washed out and the system is destabilized. This is illustrated in figure~\ref{fig:modulus1inf}. The constraint can also be expressed in the form
\begin{equation}
 H \lesssim m_{3/2}\,,
\end{equation}
where we employed $V=3\,H^2$. The constraint remains qualitatively unchanged if we couple a different inflation sector to the modulus.\footnote{See~\cite{He:2010uk} for a possible exception.}

\subsection{Two or More Moduli}\label{sec:twoormore}

In the previous example, the single modulus $T$ is apparently the field which sets the overall volume of the manifold. Destabilization of $T$, which occurs at $H\sim m_{3/2}$, triggers unacceptable decompactification of the extra dimensions. However, once we extend our consideration to multiple fields, the modulus participating in supersymmetry breaking and the modulus controlling the overall volume can generically be distinct. Consider a simple two-modulus example for which the volume is determined as
\begin{equation}\label{eq:volume}
 \mathcal{V} = (\re T_1)^{a_1/3}\,(\re T_2)^{a_2/3}\,.
\end{equation}
The scalar potential (before including the inflaton sector) shall have a minimum at $(T_{1,0},\,T_{2,0})$. At the minimum, we may then define the overall volume modulus 
\begin{equation}\label{eq:volmod}
 T_{\mathcal{V}} = a_1 \frac{T_1}{T_{1,0}} + a_2 \frac{T_2}{T_{2,0}}\,,
\end{equation}
such that for an infinitesimal change of the volume $d\mathcal{V}\propto dT_{\mathcal{V}}$. Let us assume $T_{\mathcal{V}}$ receives a large supersymmetric mass and decouples from the low energy theory. The orthogonal linear combination shall be identified with the light modulus which is stabilized by supersymmetry breaking. It becomes clear immediately that in this setup the bound $H < m_{3/2}$ cannot hold. The overall volume remains fixed as long as the inflationary energy density does not exceed the stabilization scale of the heavy volume modulus. Since the latter does not relate to supersymmetry breaking, large hierarchies between $H$ and $m_{3/2}$ can in principle be realized.\footnote{The idea of trapping a light modulus through a heavy modulus during inflation has also been applied in~\cite{Kappl:2015pxa}.}

In reality, the heavy modulus which protects the extra dimensions does not need to coincide with the volume modulus. One can easily show that $\mathcal{V}$ in~\eqref{eq:volume} remains finite given that an arbitrary linear combination $T_1 + \alpha T_2$ with $\alpha >0$ is fixed. If the heavy linear combination is misaligned with the volume modulus, the light modulus still remains protected, but receives a shift during inflation.

\begin{figure}[htp]
\begin{center}   
 \includegraphics[height=6.8cm]{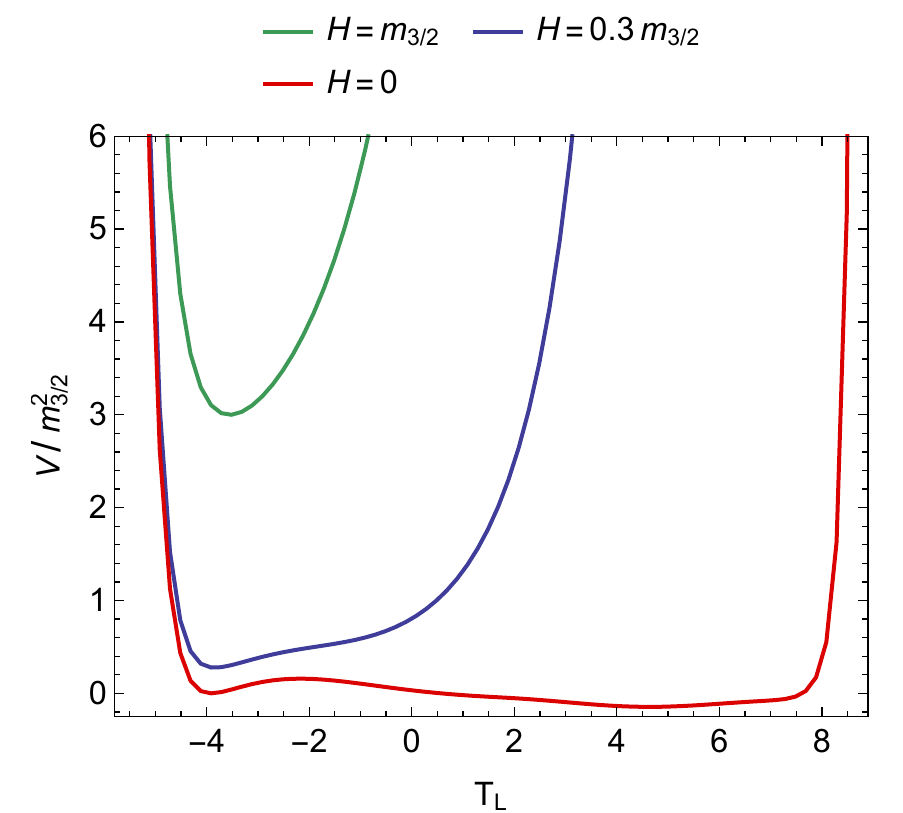}\hspace{7mm}
 \includegraphics[height=7cm]{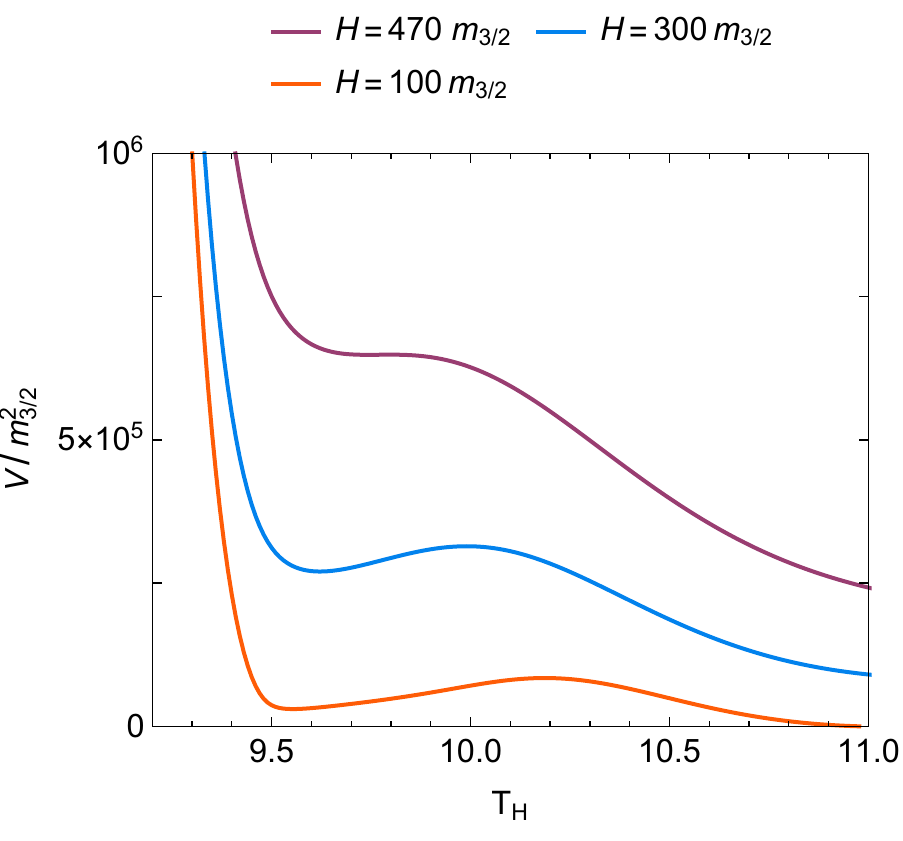} 
\end{center}
\caption{Scalar potential during inflation in the light modulus (left panel) and heavy modulus direction (right panel). For each $\Thlcaption$ and $H$, the remaining fields have been set to their corresponding minima.}
\label{fig:mod2inf}
\end{figure}

\begin{figure}[htp]
\begin{center}   
 \includegraphics[width=10cm]{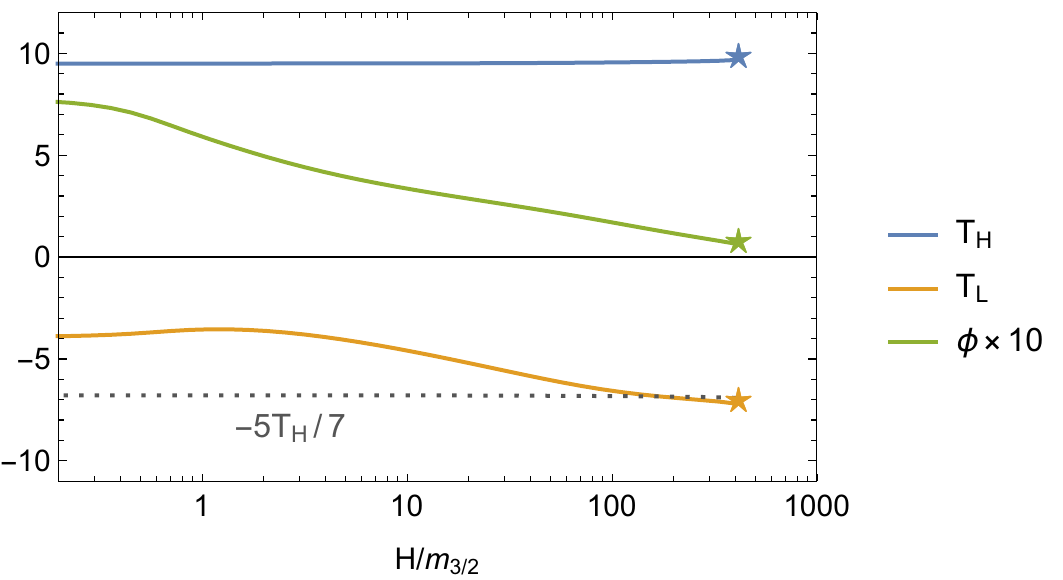}
\end{center}
\caption{Minima of $\Thcaption$, $\Tlcaption$, $\phi$ as a function of the Hubble scale. Moduli destabilization occurs at $H\simeq 470 m_{3/2}$ as indicated by the stars.}
\label{fig:vevs}
\end{figure}

In order to be more explicit, let us consider the two-modulus example of section~\ref{sec:severalmoduli}. We add the inflation sector again imposing~\eqref{eq:simple_inflation}. The scalar potential along the inflationary trajectory is
\begin{equation}\label{eq:2modinf}
 V  = V_{\text{mod}} +  \frac{e^{|\phi|^2}\ }{(\overline{T}_1+T_1)(\overline{T}_2+T_2)^6}\, W^{\alpha}W_{\alpha}\,.
\end{equation}
Inflation tends to destabilize moduli since the potential energy is minimized at $T_{1,2}\rightarrow\infty$. However, the direction $\Th=T_1+T_2$ is protected by the heavy modulus mass $m_{\Th}$. As long as $H\ll m_{\Th}$, the heavy modulus remains close to its vacuum expectation value. For fixed $\Th$, the inflaton potential energy term (second term on the right-hand side of~\eqref{eq:2modinf}) is minimized at
\begin{equation}\label{eqref:infminimum}
 \Tl = -\frac{5}{7}\, \Th\,.
\end{equation}
Hence, $\Tl$ remains protected as long as $\Th$ is stabilized. It, nevertheless, receives a shift during inflation since $\Th$ is not exactly aligned with the volume modulus. In the left panel of figure~\ref{fig:mod2inf}, we depict the scalar potential in the light modulus direction for different choices of $H$. For each value of $\Tl$ and $H$, we have minimized the potential in meson and heavy modulus direction.
It can be seen that the light modulus remains stabilized even for $H> m_{3/2}$. With growing $H$ it becomes heavier due to the Hubble mass term induced by inflation. This holds as long as the heavy modulus is not pushed over its potential barrier. For our numerical example, destabilization of the heavy modulus occurs at $H\simeq 470\,m_{3/2}$ as can be seen in the right panel of the same figure. The minima of $\Th$, $\Tl$, $\phi$ as a function of the Hubble scale are depicted in figure~\ref{fig:vevs} up to the destabilization point. It can be seen that $\Tl$ slowly shifts from $\Tlmin$ to the field value maximizing the volume as given in~\eqref{eqref:infminimum}.
Our findings can easily be generalized to systems with many moduli. In this case, an arbitrary number of light moduli remains stabilized during inflation, given at least one heavy modulus ($m_{\Th}\gg H$) which bounds the overall volume. 

A particularly appealing possibility is that the modulus which protects the extra dimensions is itself the inflaton. In particular, it would seem very natural to identify the inflaton with the overall volume modulus. We will prove in the next section that this simple picture is also favored by the K\"ahler geometry of the $G_2$ manifold. Indeed, we will show that inflationary solutions only arise in moduli directions closely aligned with the overall volume modulus.

\section{Modular Inflation in M-theory}\label{sec:modularinflation}

So far we have discussed modulus stabilization during inflation without specifying the inflaton sector. In this section, we will select a modulus as the inflaton. The resulting scheme falls into the class of `inflection point inflation' which we will briefly review. We will then identify the overall volume modulus (or a closely aligned direction) as the inflaton by means of K\"ahler geometry, before finally introducing explicit realizations of inflation and moduli stabilization.

\subsection{Inflection Point Inflation}\label{sec:inflectioninflation}

Observations of the cosmic microwave background (CMB) suggest an epoch of slow roll inflation in the very early universe. The nearly scale-invariant spectrum of density perturbations sets constraints on the first and second derivative of the inflaton potential 
\begin{equation}
 |V'|,\,|V''|\ll V\,.
\end{equation}
Unless the inflaton undergoes trans-Planckian excursions, the above conditions imply a nearly vanishing slope and curvature of the potential at the relevant field value. An obvious possibility to realize successful inflation invokes an inflection point with small slope, i.e.\ an approximate saddle point. Most features of this so-called inflection point inflation can be illustrated by choosing a simple polynomial potential
\begin{equation}\label{eq:eff_inflation}
 V= V_0 \left[1 + \frac{\delta}{\rho_0}(\rho-\rho_0) + \frac{1}{6\rho_0^3}(\rho-\rho_0)^3 \right] \;\;+ \;\;\mathcal{O}\Big((\rho-\rho_0)^4\Big)\,,
\end{equation}
where $\rho$ is the inflaton which is assumed to be canonically normalized, $\rho_0$ is the location of the inflection point. The coefficient in front of $(\rho-\rho_0)^4$ can be chosen such that the potential has a minimum with vanishing vacuum energy at the origin. Since the quartic term does not play a role during inflation, it has not been specified explicitly. The height of the inflationary plateau is set by $V_0$. The potential slow roll parameters follow as
\begin{equation}\label{eq:slowrollparameters}
 \epsilon_V = \frac{1}{2}\left(\frac{V^{\prime}}{V}\right)^2\,,\qquad \eta_V = \frac{V^{\prime\prime}}{V}\,.
\end{equation}
The number of e-folds $N$ corresponding to a certain field value can be approximated analytically,
\begin{equation}\label{eq:efolds}
 N \simeq N_{\text{max}}\left(\frac{1}{2}+ \frac{1}{\pi}\arctan\left[\frac{N_{\text{max}}(\rho-\rho_0)}{2\pi \rho_0^3}\right]\right) \,,\qquad N_{\text{max}} =\frac{\sqrt{2}\pi\,\rho_0^2}{\sqrt{\delta}}\,,
\end{equation}
where $N_{\text{max}}$ denotes the maximal e-fold number. Since we assume $\rho_0$ to be sub-Planckian, the slope parameter $\delta$ must be strongly suppressed for inflation to last 60 e-folds or longer. The CMB observables, namely the normalization of the scalar power spectrum $A_s$, the spectral index of scalar perturbations $n_s$ and the tensor-to-scalar ratio $r$ are determined by the standard expressions
\begin{equation}\label{eq:cmbobservables}
 A_s \simeq \frac{V}{24\pi^2\epsilon_V}\,,\quad n_s \simeq 1 - 6\,\epsilon_V + 2\,\eta_V\,,\quad   r\simeq 16\epsilon_V\,.
\end{equation}
For comparison with observation, these quantities must be evaluated at the field value for which the scales relevant to the CMB cross the horizon, i.e.\ at $N=50-60$ according to~\eqref{eq:efolds}. We can use the Planck measured values $n_s=0.96-0.97$, $A_s\simeq 2.1\times 10^{-9}$~\cite{Akrami:2018odb} to fix two parameters of the inflaton potential. This allows us to predict the tensor-to-scalar ratio
\begin{equation}\label{eq:tensor}
 r \sim \left(\frac{\rho_0}{0.1}\right)^6\times 10^{-11}\,.
\end{equation}
Inflation models rather generically require some degree of fine-tuning. This is also the case for inflection point inflation and manifests in the (accidental) strong suppression of the slope at the inflection point. In addition, the slow roll analysis only holds for the range of initial conditions which enable the inflaton to dissipate (most of) its kinetic energy before the last 60 e-folds of inflation. While initial conditions cannot meaningfully be addressed in the effective description~\eqref{eq:eff_inflation}, we note that the problem gets ameliorated if the inflationary plateau spans a seizable distance in field space. This favors large $\rho_0$ as is, indeed, expected for a modulus field. In this case, the typical distance between the minimum of the potential and an inflection point relates to the Planck scale (although $\rho_0 \lesssim 1$ to avoid uncontrollable corrections to the setup). Setting $\rho_0$ to a few tens of $M_P$, we expect $r\sim 10^{-8}-10^{-6} $ according to~\eqref{eq:tensor}. The maximal number of e-folds is $N_{\text{max}}=100-200$. While the modulus potential differs somewhat from~\eqref{eq:eff_inflation} (e.g.\ due to non-canonical kinetic terms), we will still find similar values of $r$ in the M-theory examples of the next sections.

\subsection{Identifying the Inflaton}

We now want to realize inflation with a modulus field as inflaton. Viable inflaton candidates shall be identified by means of K\"ahler geometry. This will allow us to derive some powerful constraints on the nature of the inflaton without restricting to any particular superpotential. 

Inflationary solutions feature nearly vanishing slope and curvature of the inflaton potential in some direction of field space. To very good approximation we can neglect the tiny slope and apply the supergravity formalism for stationary points (see section~\ref{sec:ConstraintsdS}). All field directions orthogonal to the inflaton must be stabilized. Hence, the modulus mass matrix during inflation should at most have one negative eigenvalue corresponding to the inflaton mass. The latter must, however be strongly suppressed against $V$ due to the nearly scale invariant spectrum of scalar perturbations caused by inflation. We can hence neglect it against the last term in~\eqref{eq:Vibarj} and require the mass matrix to be positive semi-definite. This leads to the same necessary condition as for the realization of de Sitter vacua, namely that $V_{i\bar{j}}$ must be positive semi-definite. During inflation, we expect the potential energy to be dominated by $F^{\rho}$. The curvature scalar of the one-dimensional submanifold associated with the inflaton $\rho$ (cf.~\eqref{eq:curvaturescalar}) should, hence, fulfill condition~\eqref{eq:curvaturecondition0}. The latter can be rewritten as
\begin{equation}\label{eq:Kahlercurvature}
 R_\rho^{-1} > \frac{3}{2} +\frac{3}{2}\left(\frac{H}{m_{3/2}^I}\right)^2\,.
\end{equation}
Here we introduced the inflationary Hubble scale through the relation $H=\sqrt{V/3}$ and the `gravitino mass during inflation' $m_{3/2}^I= e^{G/2}$. Note that $m_{3/2}^I$ is evaluated close to the inflection point. It is generically different from the gravitino mass in the vacuum which we denoted by $m_{3/2}$.
We notice that field directions with a small K\"ahler curvature scalar are most promising for realizing inflation. For a simple logarithmic K\"ahler potential $K = -a\log(\overline{\rho}+\rho)$, one finds $R_\rho= 2/a$. Condition~\eqref{eq:Kahlercurvature} then imposes at least $a>3$. However, more generically, we expect $\rho$ to be a linear combination of the moduli $T_i$ appearing in the $G_2$ K\"ahler potential~\eqref{eq:MKahler}. We perform the following field redefinition
\begin{equation}
\rho_i = \sum\limits_j O_{ij} \;\frac{\sqrt{a_j}}{2\,T^I_{j}} \;T_j\,.
\end{equation}
Here $T^I_{j}$ denotes the field value of $T_j$ during inflation (more precisely, at the quasi-stationary point). Without loss of generality, we assume that $T^I_{j}$ is real.\footnote{Imaginary parts of $T^I_{j}$ can be absorbed by shifting $T_{j}$ along the imaginary axis which leaves the K\"ahler potential invariant.} The matrix $O$ is an element of SO($M$), where $M$ denotes the number of moduli. The coefficients $a_i$ must again sum to $7$ for $G_2$. The above field redefinition leads to canonically normalized $\rho_i$ at the stationary point. We now choose $\rho_1\equiv \rho$ to be the inflaton and abbreviate $O_{1i}$ by $O_{i}$. The curvature scalar can then be expressed as
\begin{equation}\label{eq:g2curvature}
R_\rho= \sum\limits_i\frac{6\,O_{i}^4}{a_i} - \sum\limits_{i,j}\frac{4\,O_{i}^3\,O_{j}^3}{\sqrt{a_i a_j}}\,.
\end{equation}
Since successful inflation singles out field directions with small curvature scalar, it is instructive to identify the linear combination of moduli with minimal $R_\rho$. The latter is obtained by minimizing $R_\rho$ with respect to the $O_i$ which yields $O_i=\sqrt{a_i/7}$ and,
\begin{equation}\label{eq:bestdirection}
 \rho\propto \sum\limits_i \frac{a_i}{T_i^I}\, T_i\,.
\end{equation}
By comparison with~\eqref{eq:volmod}, we can identify this combination as the overall volume modulus (defined at the field location of inflation). The corresponding minimal value of $R_\rho= 2/7$. 

Hence, inflation must take place in the direction of the overall volume modulus or a closely aligned field direction -- as was independently suggested by modulus stabilization during inflation (see section~\ref{sec:twoormore}). In order to be more explicit, we define $\theta$ as the angle\footnote{The angle $\theta$ is defined in the $M$-dimensional space spanned by the canonically normalized $T_i$. For two linear combinations of moduli $\rho_1 = \alpha_i \hat T_i$ and $\rho_2 = \beta_i \hat T_i$, it is obtained from the scalar product $\boldsymbol{\alpha}\boldsymbol{\beta}=|\boldsymbol{\alpha}| |\boldsymbol{\beta}| \cos\theta$.
Here, $\hat T_i$ denote the canonically normalized moduli $\hat T_i = (\sqrt{a_i}/T^I_{i}) \,T_i /2$.} between $\rho$ and the volume modulus $T_{\mathcal{V}}$,
\begin{equation}\label{eq:defangle}
\cos\theta = O_i\sqrt{\frac{a_i}{7}}\,.
\end{equation}
In other words, $\cos^2\theta$ is the fraction of volume modulus contained in the inflaton. The constraint on the angle depends on the properties of the manifold. However, one can derive the lower bound
\begin{equation}
 R_\rho^{-1} < \frac{7}{6} \left(1+2\cos^2\theta\right)\,,
\end{equation}
which holds for an arbitrary number of moduli and independent of the coefficients $a_i$ (only requiring that the $a_i$ sum up to 7). If we combine this constraint with~\eqref{eq:Kahlercurvature}, we find
\begin{equation}\label{eq:cosmax}
 \cos^2\theta > \frac{1}{7} + \frac{9}{14} \left(\frac{H}{m_{3/2}^I}\right)^2\,.
\end{equation}
From this condition, it may seem sufficient to have a moderate volume modulus admixture in the inflaton. However, in the absence of fine-tuning, the second term on the right hand side is not expected to be much smaller than unity. Furthermore, for any concrete set of $a_i$, a stronger bound than~\eqref{eq:cosmax} may arise. Therefore, values of $\cos\theta$ close to unity -- corresponding to near alignment between the inflaton and volume modulus -- are preferred.

Let us, finally, point out that the lower limit on the curvature scalar also implies the following bound on the Hubble scale
 \begin{equation}\label{eq:generalbound}
H <\frac{2\, m_{3/2}^I}{\sqrt{3}}\,,
 \end{equation}
which must hold for arbitrary superpotential. One may now worry that this constraint imposes either low scale inflation or high scale supersymmetry breaking. This is, however, not the case since $m_{3/2}^I$ can be much larger than the gravitino mass in the true vacuum. Indeed, if the inflaton is not identified with the lightest, but with a heavier modulus, it appears natural to have $m_{3/2}^I\gg m_{3/2}$. Nevertheless,~\eqref{eq:generalbound} imposes serious restrictions on the superpotential. In order for the potential energy during inflation to be positive, while satisfying~\eqref {eq:Kahlercurvature}, one must require\footnote{We assume that the inflaton dominantly breaks supersymmetry during inflation.}
\begin{equation}\label{eq:Gconstraint}
 3 < G^{\rho}G_{\rho} < 7\,.
\end{equation}
A single instanton term $W\supset e^{-S}$ in the superpotential would induce $G^{\rho}G_{\rho}\sim S^2$. Since perturbativity requires $S\gg 1$, one typically needs to invoke a (mild) cancellation between two or more instanton terms in order to satisfy~\eqref{eq:Gconstraint}.

\subsection{An Inflation Model}
We now turn to the construction of an explicit inflation model. For the moment, we ignore supersymmetry breaking and require inflation to end in a supersymmetric Minkowski minimum. Previous considerations suggested the overall volume modulus as inflaton candidate. The simplest scenario of just one overall modulus and a superpotential generated from gaugino condensation does, however, not give rise to an inflection point with the desired properties. The minimal working example, therefore, invokes two moduli $T_{1,2}$. One linear combination $\Th$ is assumed to be stabilized supersymmetrically with a large mass $m_{\Th}\gg H$ at $\Thmin$. This is achieved through the superpotential part $\mathcal{W}(\Th)$ which could e.g.\ be of the form described in section~\ref{sec:severalmoduli}. The orthogonal, lighter linear combination $\rho$ is the inflaton. It must contain a large admixture of the overall volume modulus.

As an example, we take superpotential and K\"ahler potential to be of the form,
\begin{align}\label{eq:twomodinf}
 W &= \mathcal{W}(T_1+T_2) + \sum\limits_i A_i e^{-2\pi T_1/N_i}  \,,\nonumber\\
 K &= - a_1 \log\left(\overline{T}_1+T_1\right)-a_2 \log\left(\overline{T}_2+T_2\right) \,.
\end{align}
The heavy modulus can be defined as $\Th= (T_1 + T_2)/2$ in this case. In the limit, where $\Th$ becomes infinitely heavy, integrating out $\Th$ at the superfield level is equivalent to replacing $\Th$ by $\Thmin$ in the superpotential and K\"ahler potential, i.e.\ $T_1 \rightarrow \Thmin +\rho$ and $T_2 \rightarrow \Thmin-\rho$. We consider the case, where inflation proceeds along the real axis. The scalar potential features terms which decrease exponentially towards large $\rho$ which originate from $W$ and its derivatives. At the same time, the prefactor $e^K$ has positive slope if we choose $a_2>a_1$. For appropriate parameters, the interplay between the super- and K\"ahler potential terms leads to an inflection point suitable for inflation. 

\begin{table}[t]
\begin{center}
 \begin{tabular}{|c|ccccccccccc|}
 \hline
 PS& $a_1$ & $a_2$ & $A_1$ & $A_2$ & $A_3$ & $A_4$ & $N_1$ & $N_2$ & $N_3$ & $N_4$ & $\Thmin$\\
 \hline\hline
1& $1$ & $6$ & $1$ & $-1.18$ & $0.719766$ & $-0.178645$ & $11$ & $15$ & $19$ & $23$ & $7.8$ \\
2& $2$ & $5$ & $-1.35$ & $2.16245$ & $-0.918729$ & $-$ & $15$ & $17$ & $19$ & $-$ & $8.2$ \\
 \hline  
 \end{tabular} 
\end{center}
\caption{Input parameter sets PS$\,$1 and PS$\,$2 which give rise to the potential shown in figure~\ref{fig:infpot}. Two input parameters are specified with higher precision. This is required to (nearly) cancel the cosmological constant and to ensure that the spectral index matches precisely with observation.}
\label{tab:ps}
\end{table}

\begin{figure}[htp]
\begin{center}   
 \includegraphics[height=5.4cm]{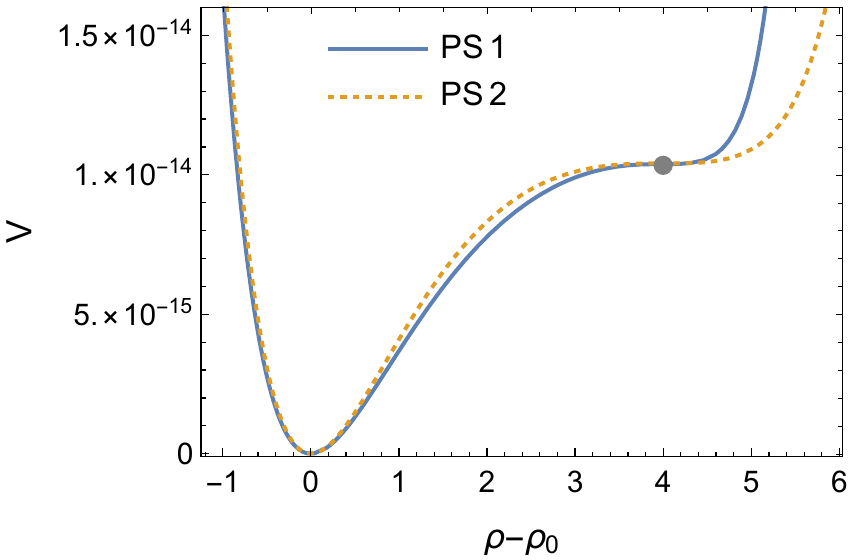}\hspace{7mm}
 \includegraphics[height=5.4cm]{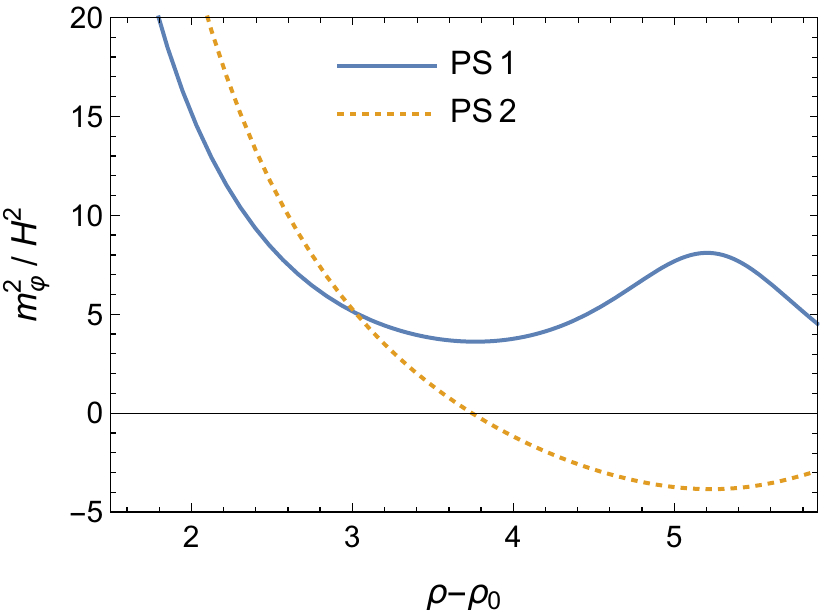} 
\end{center}
\caption{In the left panel, the inflaton potential is shown for the two parameter sets of table~\ref{tab:ps}. The inflection point at $\rho-\rho_0=4$ is indicated by the thick gray dot. In the left panel, the squared mass of the axion direction is shown in units of $H^2$.}
\label{fig:infpot}
\end{figure}

We have previously shown model-independently that the inflaton must be volume modulus-like. But how do the constraints from K\"ahler geometry actually enter the concrete setup? For this, we have to look at the axion direction $\varphi$ orthogonal to the inflaton. In table~\ref{tab:ps} we provide two parameter choices (PS$\,$1 and PS$\,$2) which give rise to a similar scalar potential along the real axis (see left panel of figure~\ref{fig:infpot}).

However, only PS$\,$1 leads to a viable inflationary scenario, while PS$\,$2 suffers from a tachyonic instability in the axion direction (at the inflationary plateau). This can be seen in the right panel of figure~\ref{fig:infpot}, where we depict the axion mass as a function of the inflaton field value. 

The reason for the tachyonic instability of PS$\,$2 becomes clear, when we study the nature of the inflaton. We express the inflaton in terms of canonically normalized moduli,
\begin{equation}\label{eq:defThat}
 \rho = O_1 \,\hat T_1 + O_2 \,\hat T_2\,,\qquad \hat T_i=\frac{\sqrt{a_i}}{2\,T^I_i} \,T_i\,.
\end{equation}
The coefficients $O_i$ determine the angle between inflaton and overall volume modulus (cf.~\eqref{eq:defangle}). In table~\ref{tab:psout} we provide the angle, the corresponding curvature scalar and the ratio $m_{3/2}^I/H$ for the two parameter sets. One can easily verify that, for PS$\,$1, the inflaton is sufficiently volume modulus-like to satisfy the constraint~\eqref{eq:cosmax} on the angle (analogously, the curvature scalar is small enough to satisfy~\eqref{eq:Kahlercurvature}). Successful inflation can, therefore, be realized. For PS$\,$2, the situation is different since the same condition is violated. The tachyonic instability which prevents inflation for PS$\,$2 is, hence, due to the misalignment between the (would-be-)inflaton and the volume modulus. 

\begin{table}
\begin{center}
 \begin{tabular}{|c|cccccc|}
 \hline
 PS& $O_1$ & $O_2$   & $\cos^2\theta$ & $ R_\rho$ & $H$ & $m_{3/2}^I$\\
 \hline\hline
 1 & $0.12$& -$0.99$ & $0.76$         & $0.34$    & $5.9\times 10^{-8}$ & $1.7\,H$\\
 2 & $0.29$& -$0.96$ & $0.42$         & $0.47$    & $5.9\times 10^{-8}$ & $1.1\,H$\\
\hline
 \end{tabular} 
\end{center}
\caption{Derived parameters for the inputs PS$\,$1 and PS$\,$2 from table~\ref{tab:ps}.}
\label{tab:psout}
\end{table}

For the parameter choice PS$\,$1, the inflationary observables can be determined from the slow roll expressions~\eqref{eq:cmbobservables}, where the normalization of the kinetic term has to be taken into account (the slow roll parameters are defined as derivatives with respect to the canonically normalized inflaton in~\eqref{eq:slowrollparameters}). The observables are consistent with present CMB bounds, specifically we find
\begin{equation}\label{eq:nsrAs}
 n_s=0.96\,,\qquad r=3\times 10^{-7}\,,\qquad A_s = 2\times 10^{-9}\,.
\end{equation}
The tensor-to-scalar ratio falls in the expected range for inflection point inflation with a modulus (see section~\ref{sec:inflectioninflation}). 

From a theoretical point, it is interesting that the inflationary plateau can be turned into a de Sitter minimum through a small parameter deformation. If we, for example, increase the value of $\Thmin$ (or change one of the $A_i$) for PS$\,$1 slightly, the potential develops a minimum close to the inflection point. The consistency of de Sitter vacua in the moduli potential follows from the $G_2$ K\"ahler potential which has a curvature scalar as small as 2/7 on the submanifold associated with the volume modulus -- in contrast to other prominent string theory constructions (see section~\ref{sec:ConstraintsdS}).

\subsection{Inflation and Supersymmetry Breaking}
In the final step, we wish to construct a more realistic model which incorporates inflation and supersymmetry breaking simultaneously. The plan is to augment the inflation sector of the previous section by the supersymmetry breaking sector comprised of the light modulus and the meson field (cf. section~\ref{sec:g2vacua}). 

The minimal example contains three moduli fields $T_{1,2,3}$ which form the linear combinations $\Th$, $\rho$ and $\Tl$. The inflaton $\rho$ must be approximately aligned with the volume modulus. An orthogonal light modulus $\Tl$ participates in supersymmetry breaking. The third modulus direction $\Th$ is stabilized supersymmetrically at a mass scale above the inflationary Hubble scale. While it does not play a dynamical role, its vacuum expectation value manifests in the K\"ahler potential of the lighter degrees of freedom. It assists in generating the plateau in the inflaton potential. The superpotential is chosen such that a mass hierarchy $m_{\Th} \gg m_{\rho} \gg m_{\Tl}$ arises in the vacuum. This can be achieved via the form
\begin{equation}
 W = \mathcal{W}(\Th) + \mathscr{W}(\Th,\rho) + w(\Th,\Tl)\,.
\end{equation}
All three superpotential parts originate from gaugino condensation. The desired mass pattern is realized through an appropriate hierarchy in the condensation scales in $\mathcal{W}$, $\mathscr{W}$ and $w$, respectively. For concreteness, we will make the following identification
\begin{equation}\label{eq:Tdefinition}
T_1 = \frac{\Th}{3} + \frac{\rho}{6} + \frac{\Tl}{2}\,,\qquad
T_2 = \frac{\Th}{3} + \frac{\rho}{6} - \frac{\Tl}{2}\,,\qquad
T_3 = \frac{\Th}{3} - \frac{\rho}{3}\,,
\end{equation}
which is just one of many possibilities. Without specifying $\mathcal{W}$ explicitly, we assume $\mathcal{W}=\WH = 0$ at $\Thmin$. As shown previously, this can e.g.\ be achieved via three gaugino condensation terms (see section~\ref{sec:severalmoduli}). In the limit of very large mass $m_{\Th}$, integrating out the heavy modulus then simply amounts to replacing $\Th$ by $\Thmin$ in the superpotential and K\"ahler potential. In addition, we choose
\begin{equation}
 w =  A_1 \,\phi^{-\frac{2}{N_1}} \, e^{-\frac{2\pi f_1}{N_1}} + A_2 \, e^{-\frac{2\pi f_2}{N_2}}\,,\qquad
 \mathscr{W} = \sum\limits_{i=3}^6 A_i e^{-\frac{2\pi f_i}{N_i}}\,.
\end{equation}
The gauge kinetic functions are defined as
\begin{equation}
 f_{1,2} = 2\,T_1 + T_3=  \Th + \Tl     \,,\qquad f_{3,4,5,6} = T_1 + T_2= \frac{2}{3}\, \Th + \frac{1}{3}\,\rho\,,
\end{equation}
such that $\mathscr{W}$ only depends on $\rho$, while $w$ only depends on $\Tl$ and $\phi$ (once $\Th$ has been integrated out). The $G_2$ K\"ahler potential,
\begin{equation}
 K = - \sum\limits_{i=1}^3 a_i \log\left(\overline{T}_i+T_i\right)\,,
\end{equation}
can be expressed in terms of $\rho$, $\Tl$ via~\eqref{eq:Tdefinition}. For an exact numerical evaluation, we choose the parameter set of table~\ref{tab:inflationparameters}. 
\begin{table}[t]
\begin{center}
 \begin{tabular}{|ccccccccccccccc|}
 \hline
 $a_1$ & $a_2$ & $a_3$ & $A_1$ & $A_2$ & $A_3$ & $A_4$ & $A_5$ & $N_1$ & $N_2$ & $N_3$ & $N_4$ & $N_5$ & $N_6$ & $T_{A,0}$\\[1mm] \hline\hline &&&&&&&&&&&&&&\\[-3mm]
 $\frac{1}{2}$ & $2$ & $\frac{9}{2}$ & $-7$ & $0.117$ & $-4.9$ &  $22.52$ & $-20.52678$ & $8$ & $10$ & $24$ & $30$ & $32$ & $38$ & $21.7$\\ \hline
 \end{tabular} 
\end{center}
\caption{Parameter choice giving rise to the inflaton potential shown in figure~\ref{fig:infpot2}. The parameter $A_5$ has been specified with higher precision in order to ensure that inflation with the correct spectral index arises. Cancellation of the cosmological constant fixes the remaining input parameter, $A_6=2.4213062895$.}
\label{tab:inflationparameters}
\end{table}

\begin{figure}[t]
\begin{center}   
 \includegraphics[height=6.5cm]{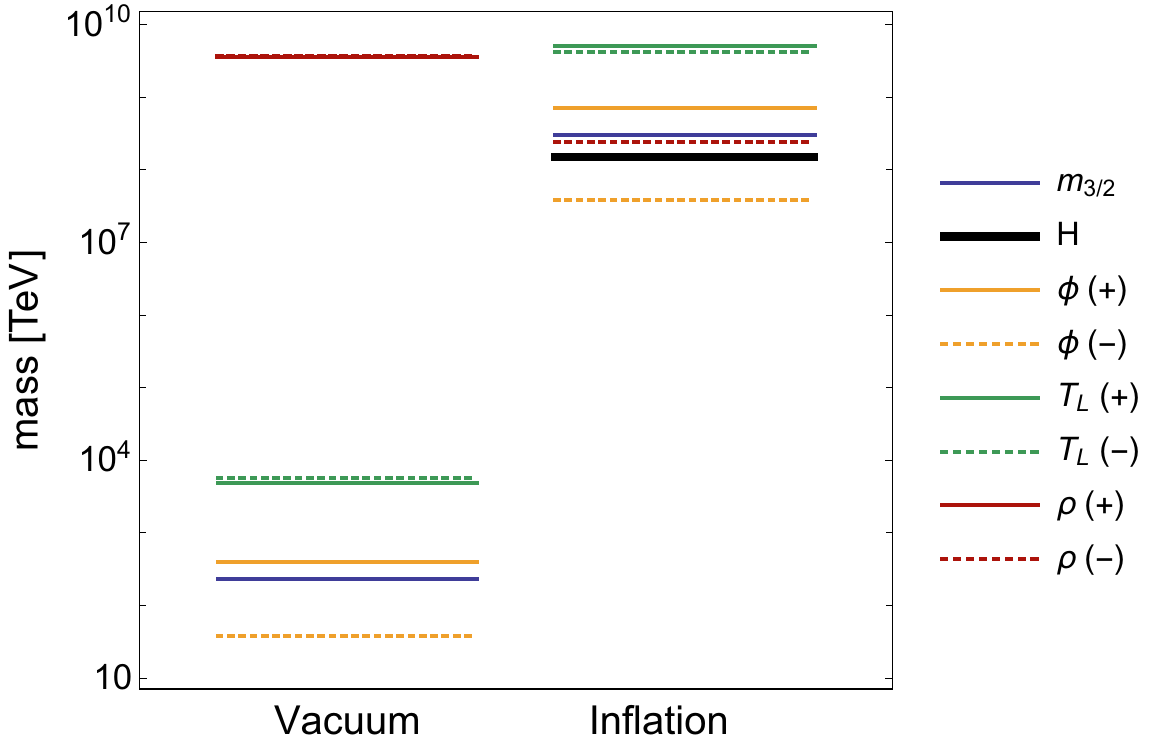}
\end{center}
\caption{Spectrum of scalar (+) and pseudoscalar (-) masses in the vacuum and during inflation. The dominant field components of the mass eigenstates are given in the plot legend (the orange lines e.g. refer to the meson-like mass eigenstates). Also shown are the gravitino mass and the Hubble parameter during inflation.}
\label{fig:spectrum}
\end{figure}

The latter gives rise to a Minkowski minimum with broken supersymmetry at $\phi_0=0.78$, $\rho_0=-3.5$, $\Tlmin=6.7$ (corresponding to $T_1=10$, $T_2=3.3$, $T_3=8.4$ in the original field basis). An additional AdS minimum appears outside the validity of the supergravity approximation ($T_2 <1$). In the Minkowski minimum, where we can trust our calculation, the mass spectrum shown in figure~\ref{fig:spectrum} arises. The light modulus and meson are responsible for supersymmetry breaking. Their masses cluster around the gravitino mass $m_{3/2}\sim 200\tev$. A slight suppression of the meson-like axion mass arises due to an approximate shift symmetry (see section~\ref{sec:severalmoduli}). The inflaton is substantially heavier compared to the other fields since it decouples from supersymmetry breaking.

\begin{figure}[t]
\begin{center}   
 \includegraphics[height=5cm]{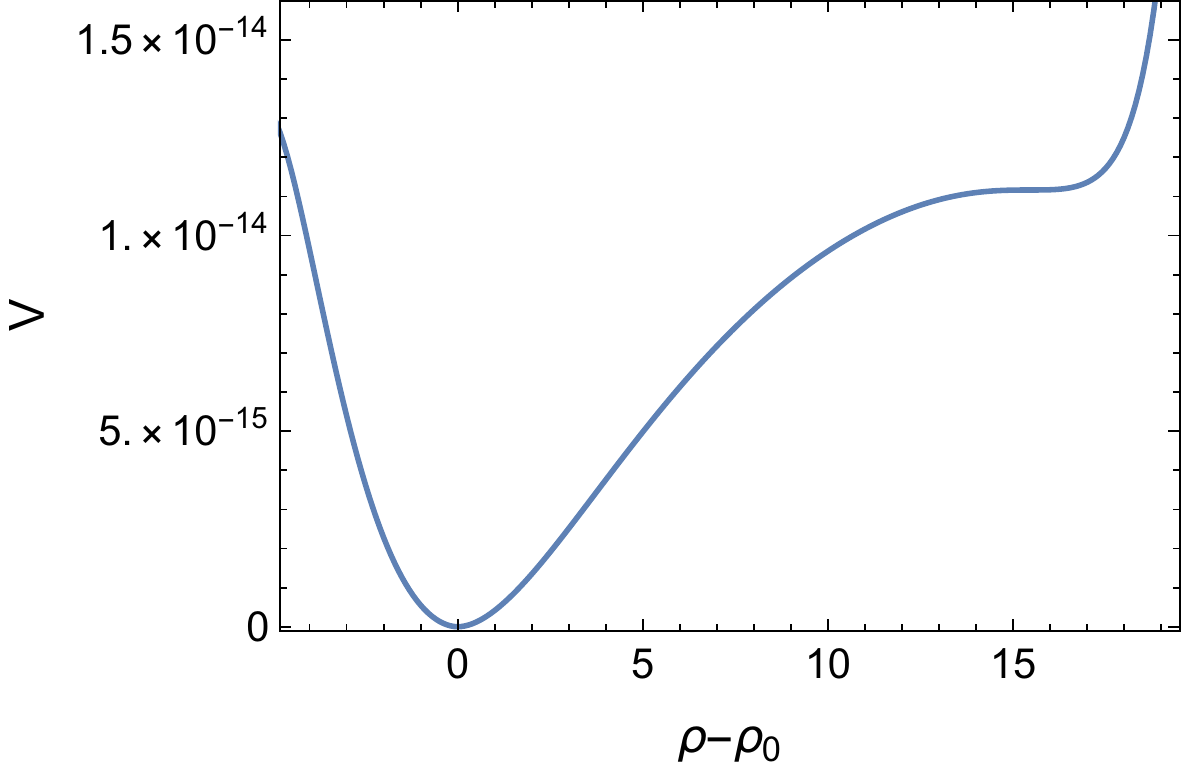}
\end{center}
\caption{Scalar potential in the inflaton direction with the other fields eliminated through their minimization condition.}
\label{fig:infpot2}
\end{figure}

Inflation occurs along the real axis of $\rho$. The potential along this direction is shown in figure~\ref{fig:infpot2}, where the remaining fields have been set to their $\rho$-dependent minima. A (quasi-stationary) inflection point occurs at $\rho-\rho_0=15.5$, where we can still trust the supergravity approximation. Corrections to the moduli K\"ahler potential, which are expected at small compactification volume, are suppressed in this regime. Even if they slightly perturbed the inflaton potential, this could easily be compensated by adjusting the superpotential parameters. Inflation, hence, appears to be robust with respect to any higher order effects.

For applying the constraints from K\"ahler geometry, we express the inflaton in terms of the (canonically normalized) original field basis
\begin{equation}
 \rho \propto 0.09\,\hat{T}_1 + 0.07 \,\hat{T}_2 - 0.99 \,\hat{T}_3\,,
\end{equation}
where the $\hat{T}_i$ have been defined in~\eqref{eq:defThat}. As can be seen, the inflaton is dominantly $T_3$. The curvature scalar along the inflaton direction is $R_\rho=0.45$. The Hubble scale and the gravitino mass during inflation are depicted in figure~\ref{fig:spectrum}. One easily verifies that the curvature constraint~\eqref{eq:Kahlercurvature} is satisfied and viable inflation without tachyons can thus be achieved. This can be related to the fact that the inflaton is sufficiently aligned with the volume modulus. The fraction of volume modulus contained in the inflaton is given by $\cos^2\theta=0.54$, in agreement with~\eqref{eq:cosmax}.

In figure~\ref{fig:spectrum}, we also provide the scalar mass spectrum during inflation. The inflaton mass is not shown since its squared mass is negative as required by the constraints on the spectral index, specifically $m_{\rho}^2=-0.05\,H^2$ during inflation (corresponding to $\eta_V=-0.015$). The other scalars receive positive Hubble scale masses during inflation (as described in section~\ref{sec:twoormore}). Only the meson-like axion is about an order of magnitude lighter than $H$ due to the approximate shift symmetry. The resulting isocurvature perturbations in the light axion are not expected to be dangerous since they are transferred into adiabatic perturbations once the axion has decayed into radiation. For the parameter example, this decay occurs before primordial nucleosynthesis (BBN). 

\begin{figure}[htp]
\begin{center}   
 \includegraphics[height=7cm]{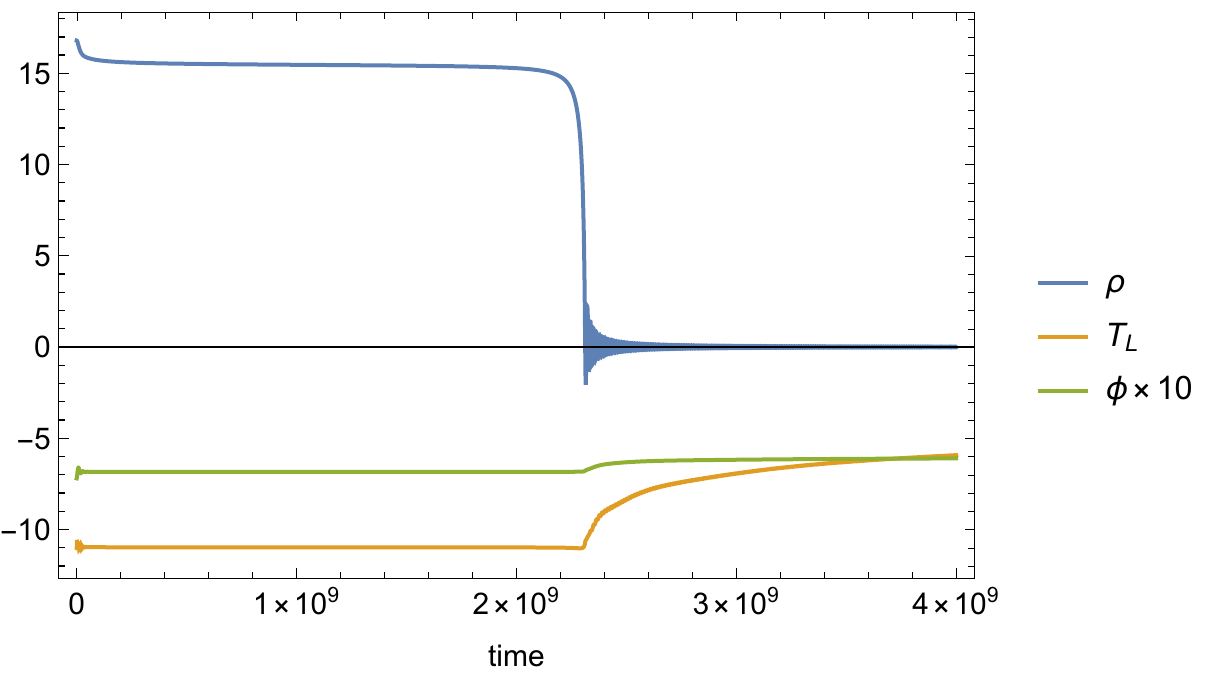}
\end{center}
\caption{Solution to the coupled system of equations of motion for the fields $\rho$, $\Tlcaption$, $\phi$.}
\label{fig:fields}
\end{figure}

In order to describe the dynamics of the multi-field system, the coupled equations of motion need to be solved. For non-canonical fields, the most general set of equations reads~\cite{Sasaki:1995aw}
\begin{equation}
\ddot{\psi}^\alpha+\Gamma^{\alpha}_{\beta\gamma}\dot{\psi}^{\beta}\dot{\psi}^{\gamma}+3H\dot{\psi}^{\alpha}+\mathcal{G}^{\alpha\beta}\frac{\partial V}{\partial\psi^{\beta}}=0\,. 
\end{equation}
Here the fields $\psi^{\alpha}$ label the real and imaginary parts of $\rho$, $\Tl$, $\phi$. The field space metric $\mathcal{G}_{\alpha\beta}$ can be determined from the K\"ahler metric and 
$\Gamma^{\alpha}_{\beta\gamma}$ is the Christoffel symbol with respect to the field metric $\mathcal{G}_{\alpha\beta}$ and its inverse $\mathcal{G}^{\alpha\beta}$. The solution to the field equations is depicted in figure~\ref{fig:fields}. For a range of initial conditions, the fields approach the inflationary attractor solution. This means that $\Tl$, $\phi$ settle at finite field-values which do not depend on the initial condition after a few oscillations. Their minima during inflation, however, differ from their vacuum expectation values. The inflaton $\rho$ slowly rolls down its potential close to the inflection point. Inflation ends when it reaches the steeper part of the potential. Then, $\rho$ oscillates around its vacuum expectation value with the amplitude decreasing due to the Hubble friction. The inflationary observables can again be determined from a slow roll analysis. The parametric example was chosen to be consistent with observation. It has
\begin{equation}
 n_s=0.97\,,\qquad r=5\times 10^{-7}\,,\qquad A_s = 2\times 10^{-9}\,.
\end{equation}
The field evolution shown in figure~\ref{fig:fields} spans five orders of magnitude in energy. All scalar fields remain stabilized over the full energy range. After inflation, the volume of the compactified manifold remains protected by the large inflaton mass. If the scalar potential features more than one minimum, the post-inflationary field evolution should ensure that the universe ends up in the desired vacuum.\footnote{In the parameter example, an additional AdS minimum occurs. It may, however, get lifted since it appears outside the range, where we can trust the supergravity calculation.} This might impose additional constraints on the moduli couplings including those to the visible sector. A comprehensive discussion of the reheating process is, however, beyond the scope of this work. Let us just note that the energy density stored in the light degrees of freedom redshifts slower than the thermal energy of the radiation bath and may dominate the energy content of the universe before they decay. We, therefore, expect a non-standard cosmology with late time entropy production to occur (see~\cite{Acharya:2008bk}). Notice that this scenario is consistent with the observed element abundances since all particles are sufficiently heavy to decay before BBN.

\newpage

\section{Conclusion}

M-theory compactified on a manifold of $G_2$ holonomy successfully describes many microphysical features of our world. It has chiral fermions interacting via gauge forces and explains the hierarchy of scales. We have now identified the inflaton within this theory. The latter is essentially the overall volume modulus of the compactified region (or a closely aligned field direction). This statement is model-independent and derives from the K\"ahler geometry of the $G_2$ manifold.

We provided concrete realizations of volume modulus inflation which satisfy all consistency conditions. Inflation occurs close to an inflection point in the scalar potential. In the relevant parameter regime, string theory corrections to the supergravity approximation are under full control. We solved the system of coupled field equations and proved that all moduli are stabilized during inflation. The scalar fields orthogonal to the inflaton receive Hubble mass terms such that inflation is effectively described as a single field slow roll model. However, several scalar fields are displaced from their vacuum expectation values during inflation. They are expected to undergo coherent oscillations when the Hubble scale drops below their mass. The energy stored in these degrees of freedom generically induces late time entropy production at their decay (which happens before BBN). 

The scale of inflation emerges from hidden sector strong dynamics. The Planck scale is the only dimensionful input to the theory. We predict $V^{1/4}\sim 10^{15}\gev$ and the corresponding tensor-to-scalar ratio $r\sim 10^{-6}$. Despite the large energy density of inflation, the theory is consistent with, and generically has low energy supersymmetry. It has a de Sitter vacuum in which the (s)goldstino dominantly descends from a hidden sector meson field. Supersymmetry breaking is transmitted to the visible sector via gravity mediation. It generates a hierarchy with heavy sfermions and lighter gauginos. The gauginos are expected to reside at the TeV scale, close to the present LHC sensitivity.

While experiments will not directly probe the inflaton of compactified M-theory, indirect evidence may be collected. This is because inflation sets the initial conditions for a non-thermal cosmology which affects many other phenomena including baryogenesis and dark matter. Further predictions of the compactified M-theory will soon be tested by laboratory experiments.

\section*{Acknowledgments}
We would like to thank Scott Watson for helpful comments on the manuscript. MW acknowledges support by the Vetenskapsr\r{a}det (Swedish Research Council) through contract No. 638-2013-8993 and the Oskar Klein Centre for Cosmoparticle Physics and the LCTP at the University of Michigan, and both of us from DoE grant DE-SC0007859.

\bibliography{moduli}
\bibliographystyle{stylefile}

\end{document}